# MARADONER: Motif Activity Response Analysis Done Right

Georgy Meshcheryakov[1] Andrey I. Buyan[1]
iam@georgy.top andreybuyanchik@gmail.com

**Abstract**

Inferring the activities of transcription factors from high-throughput transcriptomic or open chromatin profiling, such as RNA-/CAGE-/ATAC-Seq, is a long-standing challenge in systems biology. Identification of highly active master regulators enables mechanistic interpretation of differential gene expression, chromatin state changes, or perturbation responses across conditions, cell types, and diseases. Here, we describe MARADONER, a statistical framework and its software implementation for motif activity response analysis (MARA), utilizing the sequence-level features obtained with pattern matching (motif scanning) of individual promoters and promoter- or gene-level activity or expression estimates. Compared to the classic MARA, MARADONER (MARA-done-right) employs an unbiased variance parameter estimation and a bias-adjusted likelihood estimation of fixed effects, thereby enhancing goodness-of-fit and the accuracy of activity estimation. Further, MARADONER is capable of accounting for heteroscedasticity of motif scores and activity estimates.

Code availability: https://github.com/autosome-ru/maradoner

## 1. Introduction

The regulation of gene expression is a fundamental process. On the transcriptional level, it is orchestrated primarily by a special class of proteins, the transcription factors (TFs), which recognize specific DNA motifs and bind to motif occurrences in gene regulatory regions, promoters and enhancers, to activate or repress transcription. Despite significant advances in high-throughput mapping of regulatory interactions, the knowledge of genome-wide cell type-specific regulatory networks in eukaryotes including humans remains incomplete.

Direct measurement of TF expression levels is often impractical due to post-translational modifications, variable protein stability, and diverse protein localization that decouple TF abundance from its functional activity. A popular workaround is to focus on TF-recognized sequence motifs and infer activities of individual motifs as proxies for respective group of TFs sharing similar DNA binding specificity. I.e., computational methods can infer motif activities by modeling downstream effects, such as gene expression variations, based on predicted binding sites [1].

A traditional approach to identification of key regulatory motifs, such as matrixREDUCE [2], FIRE [3] or HOMER [4], involves differential motif enrichment analysis e.g. comparing the sequences of regulatory regions controlling differentially expressed genes. An orthogonal approach is Motif Activity Response Analysis (MARA) [5] and its derivative ISMARA [6], which model gene expression as a linear function of motif occurrence counts in promoters, inferring the respective TF activities across samples.

Inferring motif activities has broad applications in systems biology. It facilitates the reconstruction of active regulatory networks, identification of master regulators in development or disease, and mechanistic interpretation of differential expression or chromatin responses. For example, these approaches have elucidated key regulators in cellular differentiation [5], immune responses [7], and cancer progression [8], aiding hypothesis generation for experimental validation.

ISMARA has been extended to incorporate miRNA regulation and chromatin data, demonstrating utility in diverse mammalian systems [9], [10], [11]. Newer tools, such as decoupleR [12] and IMAGE [7], use ensemble methods to filter out unlikely regulons/TFs beforehand or leverage ChIP-Seq data and the information on enhancers to augment a motif scores, but underneath stick to the MARA-like linear model to estimate motif activities. These approaches have improved scalability – a necessity as the volume of genomic sequence data continues to expand across diverse biological domains [13] – allowing for application to single cell data or offer improved accuracy, but still suffer from biases in variance estimation, inefficient fixed-effect estimators (e.g., ordinary least squares instead of generalized least squares), behaving suboptimally in small-sample scenarios, and making questionable assumptions like zero mean activities that obscure repressor-activator distinctions.

[1]Institute of Protein Research, Russian Academy of Sciences

Here, we introduce MARADONER (Motif Activity Response Analysis Done Right), a novel statistical framework reinventing MARA with unbiased estimation of regulatory motif activities from the promoter-level gene expression data. Building on the linear modeling paradigm of MARA, MARADONER employs orthogonal transformations and restricted maximum likelihood to yield unbiased variance estimates and best linear unbiased estimators (BLUE) of fixed effects. It explicitly models motif-specific variances, enhancing goodness-of-fit particularly the case of small number of samples, and optionally accounts for the promoter-specific heteroscedasticity for greater flexibility. Available as open-source software, MARADONER advances regulatory inference by providing robust, interpretable insights into gene networks. The outline of the paper is as follows:

1. In Section 2,we explain notation and introduce necessary identities from linear algebra along with the formal notation;
2. In Section 3, we introduce the MARADONER model;
3. In Section 4, we explain the algorithm behind MARADONER that makes estimating the model's parameters a feasible problem.
4. In Section 5, we introduce an enhanced version of the MARADONER model, allowing for a greater heteroscedasticity via including gene-specific (or promoter-specific) variances to the model;
5. In Section 6, we discuss techniques to cluster motifs, which could be useful to decrease the number of model parameters; ;
6. In Section 7 the maximum-a-posteriori estimator of motif activities is described;
7. In Section 8 we describe statistical tests present in the MARADONER based on asymptotic properties of estimates and the posterior distribution of motif activities;
8. Section 9 provides a short overview of the MARADONER's algorithm;
9. Section 10 explains an extension of the MARADONER framework to build gene regulatory networks (GRNs), that is, to identify statistically significant TF-gene pairs for which a TF has a positive impact on explaining a gene's variance;
10. Section 11 provides an evaluation of the MARADONER methods using synthetic data and a comparison of MARADONE with MARA.

# 2. Preliminary info

## 2.1. Notation

Matrices will be denoted by capital letters, while lowercase letters will be used for scalars. Vectors (matrices of size $n \times 1$) are also denoted in lowercase, but bold font is used. The identity matrix of size $n$ will be denoted by $I_n$. $O_n$ is defined as a square matrix of ones, with dimensions of n by n. The $J_\Xi$ matrix is defined as a zero matrix with diagonal elements, whose positions are listed in a set of integers ξ, set to $1$ (alternatively, $\Xi$ can be an integer value $i$, then $J_i$ is just a zero matrix with $i$-th diagonal element set to $1$). The symbol $\mathbf{1}_n$ denotes a vector of $n$ ones. The symbol $\boldsymbol{e}_k^n$ denotes the $k$ basis vector in the $n$-dimensional space. In certain instances, $\boldsymbol{e}_k^n$ will be written simply as $\boldsymbol{e}_k$ if the dimensionality can be readily ascertained from the context. $\otimes$ is a Kronecker product. $\odot$ is a Hadamard (element-wise) product. Matrix multiplication symbol $\times$ is usually omitted with a notable exception of line breaks. $\lfloor x \rfloor$ is a floor function, $\{x\}$ is a fractional part of $x$, $x\%y = x - y\left\lfloor \frac{x}{y} \right\rfloor$.

$\texttt{vec}[X]$ is a vectorization of matrix $X$. $\texttt{mat}_n^m[\boldsymbol{x}]$ is a matricization of vector $\boldsymbol{x}$ of length $nm$, turning it into a matrix of size $n \times m$ in a fashion such that it is an inverse of the $\texttt{vec}$ operator. $\texttt{diag}[X]$ is either a vector of the diagonal of $X$ if $X$ is a square matrix, else if $X$ is a vector, then it constructs a diagonal matrix out of it.

For the most part, $1$-based indexing is used, except for the Section 4.6.4.3, where $0$-based indexing is used instead for brevity.

$\mathcal{MN}$ is a matrix-variate normal distribution [14]. This is merely a convenient shorthand notation for a particular instance of a multivariate normal distribution:

$$X \sim \mathcal{MN}(M, A, B) \leftrightarrow \texttt{vec}[X] \sim \mathcal{N}(\texttt{vec}[m], B \otimes A). \tag{1}$$

## 2.2. Some useful identities

In subsequent sections, we shall frequently employ well-known identities listed below. Let $A, B, C, D, X$ be matrices of appropriate shapes. Then the following identities hold true:

- Push-through identity:

$$(AA^\top + I)A = A(A^\top A + I) \leftrightarrow A(A^\top A + I)^{-1} = (AA^\top + I)^{-1}A. \tag{2}$$

- Woodbury identity:

$$\left(AA^\top + \sigma I_p\right)^{-1} = \frac{1}{\sigma}I_p - \frac{1}{\sigma}A(\sigma I_m + A^\top A)^{-1}A^\top. \tag{3}$$

- Determinant lemma:

$$\det\left(\sigma I_p + AA^\top\right) = \det\left(I_m + \frac{1}{\sigma}A^\top A\right)\sigma^p = \det(\sigma I_m + A^\top A)\sigma^{p-m} \tag{4}$$

- Derivative of the $\ln\det$ function:

$$\frac{\partial \ln\det(A)}{\partial x} = \frac{\partial\, \mathrm{tr}\ln(A)}{\partial x} = \mathrm{tr}\left\{\frac{\partial \ln(A)}{\partial x}\right\} = \mathrm{tr}\left\{A^{-1}\frac{\partial A}{\partial x}\right\}. \tag{5}$$

- Derivative of the inverse matrix:

$$\begin{aligned}\frac{\partial A^{-1}}{\partial x} = \lim_{\Delta x\to 0}\frac{(A+\Delta A)^{-1} - A^{-1}}{\Delta x} = \lim_{\Delta x\to 0}\frac{(A+\Delta A)^{-1}AA^{-1} - (A+\Delta A)(A+\Delta A)^{-1}A^{-1}}{\Delta x} = \\ = -A^{-1}\lim_{\Delta x\to 0}\frac{\Delta A}{\Delta X}A^{-1} = -A^{-1}\frac{\partial A}{\partial x}A^{-1}.\end{aligned} \tag{6}$$

- Mixed-product property of the Kronecker product:

$$(A\otimes B)(C\otimes D) = (AC)\otimes(BD). \tag{7}$$

- Singular Value Decomposition (SVD) of the Kronecker product:

If $Q_A D_A V_A$ and $Q_B D_B V_B$ are SVDs of matrices $A$ and $B$ respectiely, then

$$A\otimes B = (Q_A D_A V_A)\otimes(Q_B D_B V_B) = (Q_A\otimes Q_B)(D_A\otimes D_B)(V_A\otimes V_B), \tag{8}$$

which follows from Equation 7.

- Link between vectorization and Kronecker product:

$$(B\otimes A)\ \mathtt{vec}[X] = \mathtt{vec}\left[AXB^\top\right] \tag{9}$$

# 3. Model

Let there be $\mathring{p}$ promoters, $g$ groups, each of $N(i)$ samples, $\sum_{i=1}^{g} N(i) = n$ and let $m$ be a number of motifs. We propose the following linear model of promoters expression $Y$:

$$\underbrace{\mathring{Y}}_{\mathring{p}\times n} = \underbrace{\mathbf{1}_{\mathring{p}}}_{p\times 1}\underbrace{\boldsymbol{\mu_n}^\top}_{1\times n} + \underbrace{\boldsymbol{\mu_p}}_{\mathring{p}\times 1}\underbrace{\mathbf{1_n}^\top}_{1\times n} + \underbrace{\mathring{B}}_{p\times m}U + \mathring{\mathrm{E}},\ \mathring{\mathrm{E}}\sim\mathcal{MN}\left(0, I_{\mathring{p}}, D\right),\ U\sim\mathcal{MN}\left(\boldsymbol{\mu_m}\mathbf{1_n}^\top, \Sigma, G\right), \tag{10}$$

where $\mathring{B}$ is a matrix of motif loadings onto promoters, $U$ is a random matrix of motif activities, $\mathring{\mathrm{E}}$ is a random matrix of errors, $D = \mathtt{diag}\left\{\sigma_{G(i)}\right\}_{i=1}^{n}$ (here, $G(i)$ returns an index of the group that the $i$-th observations belongs to) and similarly is $G = \mathtt{diag}\left\{\nu_{G(i)}\right\}_{i=1}^{n}$, $\Sigma = \mathtt{diag}\{\tau_i\}_{i=1}^{m}$ is a motif variance matrix, $\mu_n$ and $\mu_p$ a sample-wise and promoter means respectively.

Note that all groups have common $\Sigma$ and distinct $\sigma$ and $\nu$. The model can be easily extended to a "more" heteroscedastic case with $\mathring{\mathrm{E}}\sim\mathcal{MN}(0, K, D)$, where $K$ is a diagonal matrix of promoter variances. We shall consider this extension briefly in Section 5.

# 4. Parameters estimation

Utilising the model from Equation 10 without modification and estimating all the parameters in $\Sigma, G, \boldsymbol{\mu_p}, \boldsymbol{\mu_n}$ may result in computationally onerous and inefficient processes. Moreover, it is widely acknowledged that the joint estimation of the variance and mean parameters with the maximum likelihood (ML) method engenders biased variance estimates. In order to address these issues, an iterative parameter-by-parameter estimation procedure is proposed.

## 4.1. Centering data

Firstly, it is necessary to obtain a method of centring data and model from Equation 10 in order to deal with $\boldsymbol{\mu_p}$ and $\boldsymbol{\mu_s}$.

### 4.1.1. A note on MARA-like de-meaning

In the classical MARA framework [6], the dependent variable $Y$ is centered across both rows and columns. In practice, if we denote $< \mathring{Y}_{i\cdot} >$ as a sample average for a fixed $i$-th dimension, it implies that we substitute data with

$$\begin{aligned} Y_{i,j} = \mathring{Y}_{i,j} - < \mathring{Y}_{i,\cdot} > - < \mathring{Y}_{\cdot,j} > + < \mathring{Y} >, \\ B_{i,j} = \mathring{B}_{i,j} - < \mathring{B}_{\cdot,j} > . \end{aligned} \tag{11}$$

In MARA, that's all the preprocessing that is done and intercepts are not incorporated into the model. However, let's reflect on how exactly this transformation might occur and what implications it has for the model in Equation 10. Let's introduce the centering matrix:

$$C_n = I_n - \frac{1}{n} O_n, \tag{12}$$

Note that $C_n \mathbf{1_n} = \mathbf{0}_n$. That means we can hope to deal with the mean components by transforming $\mathring{Y}$ with $C_p$ and $C_{\mathring{n}}$ from the left and the right:

$$C_p \mathring{Y} C_n = Y = \cancel{C_p \mathbf{1_p} \boldsymbol{\mu_n}} + \cancel{C_p \boldsymbol{\mu_p}^\top (C_{\mathring{n}} \mathbf{1_{\mathring{n}}})^\top} + C_p \mathring{B} U C_{\mathring{n}} + C_p \mathring{\mathrm{E}} C_n = BUC_n + C_p \mathring{\mathrm{E}} C_n. \tag{13}$$

Indeed, applying formulae from Equation 11 eliminates mean components, but r.v.s $U$ and $\mathring{E}$ become transformed. More specifically, their covariances matrices are now proportional to $C_p$ and $C_{\mathring{n}}$ (as $C_k^\top C_k = C_k$ due to idempotency of $C_k$). What's worse, is that $C_k$ is of rank $k-1$, producing singular covariance matrices.

Should these facts be disregarded (as they are in MARA), there is a risk of obtaining biased estimates of variances and the random effects matrix $U$. However, it can be argued that centering $\mathring{Y}$ is effectively estimating fixed effects via Ordinary Least Squares (OLS) method. This is both an unbiased and consistent estimator, yet it is certainly not an efficient one (the efficient one being the Generalized Least Squares (GLS) estimator) as it ignores the weighting imposed by the covariance structure of the model. In principle, this should not be a concern for large datasets, given the consistency of the OLS (Ordinary Least Squares) estimator.

One more point to address in regards to the original MARA model is that authors assume expected value of $U$ to be zero. That seemingly should not matter as multiplying $Y$ by the $C_n$ from the right eliminates the $\boldsymbol{\mu_m} \mathbf{1_n}^\top$ component. This, however, means that in MARA we deal with the deviations of motif activities from their mean values rather than with the motif activities per se. This lack of clarity introduces confusion and the lack of $\boldsymbol{\mu_m}$ estimates limits our capacity to differentiate between repressors and activators.

### 4.1.2. Orthogonal complement

Instead of using centering matrices $C_k$, we propose to use orthogonal complements to $\mathbf{1_k}$, $H_k$ matrix of size $(k-1) \times k$. Although it lacks the intuitive interpretation of the "de-meaning" approach from Section 4.1.1, it allows us to use the same model from Equation 10 and obtain unbiased estimates of variance parameters:

$$\begin{aligned} H_{\mathring{p}} \mathring{Y} H_n^\top = \underbrace{Y}_{\underbrace{p}_{\mathring{p}-1} \times n} H_n = \cancel{H_{\mathring{p}} \mathbf{1_{\mathring{p}}} \boldsymbol{\mu_n}} + \cancel{H_{\mathring{p}} \boldsymbol{\mu_p} \mathbf{1_n}^\top} H_n + H_{\mathring{p}} \mathring{B} U H_n + H_{\mathring{p}} \mathring{\mathrm{E}} H_n = B\, U H_n + H_{\mathring{p}} \mathrm{E} H_n = \\ = BUH_n + \mathrm{E} H_n,\ U \sim \mathcal{MN}(0, \Sigma, G),\ \mathrm{E} \sim \mathcal{MN}\left(0, \underbrace{I_p}_{H_{\mathring{p}} H_{\mathring{p}}^\top}, D\right). \end{aligned} \tag{14}$$

Here, we denoted $H_{\mathring{p}} \mathring{Y}$, $H_{\mathring{p}} \mathring{B}$, $H_{\mathring{p}} \mathring{\mathrm{E}}$ as $Y$, $B$ and $E$ respectively. If we further continue to aggregate terms, then

$$Y H_n = \underbrace{\hat{Y}}_{p \times \underbrace{\hat{n}}_{n-1}} = B\hat{U} + \hat{E},\ \hat{U} \sim \mathcal{MN}(0, \Sigma, H_n G H_n^\top),\ \hat{\mathrm{E}} \sim \mathcal{MN}\left(0, \underbrace{I_{p-1}}_{H_p H_p^\top}, H_n D H_n^\top\right). \tag{15}$$

We construct the orthogonal complement $H_k = \left[\boldsymbol{f_1^k}^\top, \boldsymbol{f_2^k}^\top, ..., \boldsymbol{f_{\{k-1\}}^k}^\top\right]^\top$ to $\mathbf{1_k}$ as follows:

1. The first vector $\hat{\boldsymbol{f}}_{\boldsymbol{1}}^{\boldsymbol{k}} = [1, -1, 0, ..., 0, ..., 0]$. Note that $\hat{\boldsymbol{f}}_{\boldsymbol{1}}^{\boldsymbol{k}\top} 1_k = 0$;
2. The second vector $\hat{\boldsymbol{f}}_{\boldsymbol{2}}^{\boldsymbol{k}} = \left[\frac{1}{2}, \frac{1}{2}, -1, 0, ..., 0, ..., 0\right]^\top$. Note that $\hat{\boldsymbol{f}}_{\boldsymbol{2}}^{\boldsymbol{k}\top} \hat{\boldsymbol{f}}_{\boldsymbol{1}}^{\boldsymbol{k}} = 0$;

3. The $j$-th vector $\hat{\boldsymbol{f}}_{\boldsymbol{j}}^{\boldsymbol{k}} = \left[\underbrace{\frac{1}{j}, ..., \frac{1}{j}}_{j}, -1, 0, ..., 0\right]^\top$, i.e. ${\hat{\boldsymbol{f}}_{\boldsymbol{j}}^{\boldsymbol{k}}}_i = \frac{1}{j}$ if $i \le j$, ${\hat{\boldsymbol{f}}_{\boldsymbol{1}}^{\boldsymbol{k}}}_i = -1$ and $0$ otherwise;
4. Finally, vectors are normalized: $\boldsymbol{f}_{\boldsymbol{j}}^{\boldsymbol{k}} = \frac{\hat{\boldsymbol{f}}_{\boldsymbol{j}}^{\boldsymbol{k}}}{\|\hat{\boldsymbol{f}}_{\boldsymbol{j}}^{\boldsymbol{k}}\|_2}$, where $\|\hat{\boldsymbol{f}}_{\boldsymbol{j}}^{\boldsymbol{k}}\|_2 = \sqrt{1 + \sum_{i=1}^{j} \frac{1}{j^2}} = \sqrt{1 + \frac{1}{j}}$.

$$H_k = \begin{pmatrix} \frac{1}{\sqrt{2}} & -\frac{1}{\sqrt{2}} & 0 & \vdots & 0 \\ \frac{\sqrt{6}}{6} & \frac{\sqrt{6}}{6} & -\frac{\sqrt{6}}{3} & \vdots & 0 \\ \dots & \dots & \dots & \ddots & \dots \\ \frac{\sqrt{k(k-1)}}{(k-1)k} & \frac{\sqrt{k(k-1)}}{(k-1)k} & \dots & \frac{\sqrt{k(k-1)}}{(k-1)k} & -\frac{\sqrt{k(k-1)}}{k} \end{pmatrix}. \tag{16}$$

$H_k$, as defined, has to be computed only once for the largest needed $k$, $H_{\hat{k}}$ for $\hat{k} < k$ is selected as a submatrix of $H_k$. This transformation is related to the Helmert matrix [15]. Namely, the Helmert matrix is $\widehat{H_k} = \left[\hat{\boldsymbol{f}}_{\boldsymbol{1}}^{\boldsymbol{k}\top}, \hat{\boldsymbol{f}}_{\boldsymbol{2}}^{\boldsymbol{k}\top}, ..., \hat{\boldsymbol{f}}_{\{\boldsymbol{k-1}\}}^{\boldsymbol{k}\ \top}, \boldsymbol{1}_{\boldsymbol{k}}^\top\right]^\top$. More practically, for large $k$, it is best to avoid explicit construction of $H_k$ matrix (just like no one really uses centering matrices $C_k$ in practice).

There are some properties of this "centering" operator, useful to our cause.

**Property 4.1.2.1** (Semi-orthogonality): $H_k$ is a semi-orthogonal matrix:
1. It is orthogonal from the right:
$$H_k H_k^\top = I_{k-1} \tag{17}$$
2. From the left, $H_k$ can be seen as a decomposition of the centering matrix $C_k$:
$$C_k = H_k^\top H_k \tag{18}$$

*Proof*:
1. It is true by construction.
2. It is easy to see that both $C_k$ and $H_k^\top H_k$ are projection matrices. They both project to subspaces orthogonal to $\boldsymbol{1}_{\boldsymbol{k}}$. Those subspaces are the same if $\mathrm{rank}(C_k) = \mathrm{rank}(H_k^\top H_k)$. The rank of $C_k$ is $k-1$. Given that $H_k$ is a semi-orthogonal matrix, its rank is $k-1$ and due to the fact that for any matrix $A$ of rank $r$ the product $AA^\top$ keeps the same rank, $\mathrm{rank}(H_k^\top H_k) = k - 1$. Hence, both $C_k$ and $H_k^\top H_k$ project to the same subspace. Given that the projection matrices are unique, $C_k = H_k^\top H_k$.

■

**Property 4.1.2.2** (Inverse): Let $D$ be a diagonal matrix $D = \mathtt{diag}\{d_i\}_{i=1}^k$. Then,
$$H_k^\top (H_k D H_k)^{-1} H_k = \frac{1}{k} \frac{\det|D|}{\det|H_k D H_k^\top|} (K - \boldsymbol{d}^{-1}\boldsymbol{d}^{-T}) = \frac{1}{\boldsymbol{1}_{\boldsymbol{k}}^\top \boldsymbol{d}^{-1}} (K - \boldsymbol{d}^{-1}\boldsymbol{d}^{-T}), \tag{19}$$
where $\boldsymbol{d}_i^{-1} = \frac{1}{d_i}$, $K = \mathtt{diag}[\boldsymbol{d}^{-1}\boldsymbol{d}^{-T}\boldsymbol{1}_{\boldsymbol{k}}]$.

*Proof*:

Let's introduce a helper square matrix $Q$:
$$Q = \begin{pmatrix} H_k \sqrt{D} \\ \boldsymbol{1}_{\boldsymbol{k}}^\top \sqrt{D^{-1}} \end{pmatrix}. \tag{20}$$

See that
$$Q^\top (QQ^\top)^{-1} Q = I_k, \tag{21}$$
which is true because $(QQ^\top)^{-1} = Q^{-1}Q^{-T}$. Let's first expand this term:

$$
\begin{aligned}
(QQ^\top)^{-1} &= \left(\begin{pmatrix} H_k\sqrt{D} \\ \mathbf{1_k}^\top\sqrt{D^{-1}} \end{pmatrix}\begin{pmatrix}\sqrt{D}H_k^\top & \sqrt{D^{-1}}\mathbf{1_k}\end{pmatrix}\right)^{-1} = \begin{pmatrix} H_kDH_k & 0 \\ 0 & \mathbf{1_k}^\top D^{-1}\mathbf{1_k} \end{pmatrix}^{-1} = \\
&= \begin{pmatrix} (H_kDH_k)^{-1} & 0 \\ 0 & \frac{1}{\mathbf{1_k}^\top D^{-1}\mathbf{1_k}} \end{pmatrix}.
\end{aligned} \tag{22}
$$

Then,

$$
\begin{aligned}
Q(QQ^\top)^{-1}Q^\top &= \begin{pmatrix}\sqrt{D}H_k^\top & \sqrt{D^{-1}}\mathbf{1_k}\end{pmatrix}\begin{pmatrix} (H_kDH_k)^{-1} & 0 \\ 0 & \frac{1}{\mathbf{1_k}^\top D^{-1}\mathbf{1_k}} \end{pmatrix}\begin{pmatrix} H_k\sqrt{D} \\ \mathbf{1_k}^\top\sqrt{D^{-1}} \end{pmatrix} = \\
&= \sqrt{D}H_k^\top(H_kDH_k)^{-1}H_k\sqrt{D} + \frac{1}{\mathbf{1_k}^\top D^{-1}\mathbf{1_k}}\sqrt{D^{-1}}\mathbf{1_k}\mathbf{1_k}^\top\sqrt{D^{-1}} = I_k.
\end{aligned} \tag{23}
$$

By transforming the equation from the left and from the right by $\sqrt{D^{-1}}$, we get:

$$
H_k^T(H_kDH_k)^{-1}H_k = D^{-1} - \frac{1}{\mathbf{1_k}^\top D^{-1}\mathbf{1_k}}D^{-1}\mathbf{1_k}\mathbf{1_k}^\top D^{-1}. \tag{24}
$$

■

**Property 4.1.2.3** (Determinant): Let $D$ be a diagonal matrix $D = \mathtt{diag}\{d_i\}_{i=1}^k$. Then,

$$
\det|H_kDH_k^\top| = \frac{1}{k}\sum_{i=1}^k \frac{1}{d_i}\det|D|. \tag{25}
$$

*Proof*: Let's complement the semi-orthogonal matrix $H_k$ to a square orthogonal matrix $Q$ by adding an extra row $\boldsymbol{a} = \frac{1}{\sqrt{k}}\mathbf{1_k}$:

$$
Q = \begin{pmatrix} H_k \\ \boldsymbol{a}^\top \end{pmatrix}. \tag{26}
$$

As $Q$ is an orthogonal matrix,

$$
\det|QDQ^\top| = \det|D|. \tag{27}
$$

At the same time,

$$
QDQ^\top = \begin{pmatrix} H_k \\ \boldsymbol{a}^\top \end{pmatrix} D \begin{pmatrix} H_k^\top & \boldsymbol{a} \end{pmatrix} = \begin{pmatrix} H_kDH_k^\top & H_kD\boldsymbol{a} \\ \boldsymbol{a}^\top DH_k^\top & \boldsymbol{a}^\top D\boldsymbol{a} \end{pmatrix}. \tag{28}
$$

Using the formula for the determinant of a block matrix,

$$
\begin{aligned}
\det|QDQ^\top| &= \det|H_kDH_k^\top|\left(\boldsymbol{a}^\top D\boldsymbol{a} - \boldsymbol{a}^\top D\left[H_k^\top(H_kDH_k^\top)^{-1}H_k\right]D\boldsymbol{a}\right) = \\
&= \det|H_kDH_k^\top|\left(\boldsymbol{a}^\top D\boldsymbol{a} - \boldsymbol{a}^\top D\left[D^{-1} - \frac{1}{\mathbf{1_k}^\top D^{-1}\mathbf{1_k}}D^{-1}\mathbf{1_k}\mathbf{1_k}^\top D^{-1}\right]D\boldsymbol{a}\right) = \\
&= \det|H_kDH_k^\top|\left(\boldsymbol{a}^\top D\boldsymbol{a} - \boldsymbol{a}^\top\left[D - \frac{1}{\mathbf{1_k}^\top D^{-1}\mathbf{1_k}}\mathbf{1_k}\mathbf{1_k}^\top\right]\boldsymbol{a}\right) = \det(H_kDH_k^\top)\frac{1}{\mathbf{1_k}^\top D^{-1}\mathbf{1_k}}\boldsymbol{a}^\top\mathbf{1_k}\mathbf{1_k}^\top\boldsymbol{a}^\top = \\
&= \det(H_kDH_k^\top)\frac{k}{\mathbf{1_k}^\top D^{-1}\mathbf{1_k}} = \det|D| \leftrightarrow \det|H_kDH_k^\top| = \frac{1}{k}\det|D|\ \mathbf{1_k}^\top D^{-1}\mathbf{1_k} = \frac{1}{k}\sum_{i=1}^k\frac{1}{d_i}\det|D|,
\end{aligned} \tag{29}
$$

where in square brackets [] we used Property 4.1.2.2. ■

**Property 4.1.2.4** (Centering): Let $D$ be a diagonal matrix $D = \texttt{diag}\{d_i\}_{i=1}^{k}$. Then,

$$\sqrt{D}H_k^\top H_k\sqrt{D} = D - \frac{1}{k}\boldsymbol{d}^{\frac{1}{2}}\boldsymbol{d}^{\frac{T}{2}}, \tag{30}$$

where $\boldsymbol{d}_i^{\frac{1}{2}} = \sqrt{d_i}$.

*Proof*: As $H_k^\top H_k = C_k$ (Equation 18),

$$\sqrt{D}H_k^\top H_k\sqrt{D} = \sqrt{D}C_k\sqrt{D} = \sqrt{D}\left(I_k - \frac{1}{k}\boldsymbol{1_k}\boldsymbol{1_k}^\top\right)\sqrt{D} = D - \frac{1}{k}\boldsymbol{d}^{\frac{1}{2}}\boldsymbol{d}^{\frac{T}{2}}. \tag{31}$$

■

This property, although trivial, allows us to compute matrix-vector products in a linear time, greatly facilitating numerical algorithms such as eigenvalue decomposition. The same logic applies to Property 4.1.2.2 – although the cost of filling up all entries of the inverse matrix is surely $O((k-1)^2)$, but in practice we will need to compute matrix-vector products only, decreasing the complexity to linear.

Next, we'll make extensive use of the proposed orthogonal centering operator $H_k$ to alleviate parameters estimation.

## 4.2. Dealing with $\boldsymbol{\mu_n}$

Due to the orthogonal nature of the $H_k$ operator and the fact that in Equation 10 row-wise covariance matrix of $E$ is an identity matrix, we can safely omit group-wise intercepts from the model:

$$\begin{aligned} H_p\mathring{Y} = Y = \cancel{H_{\mathring{p}}\boldsymbol{1_p}\boldsymbol{\mu_n}^\top} + H_{\mathring{p}}\boldsymbol{\mu_p}\boldsymbol{1_n}^\top + H_{\mathring{p}}\mathring{B}U + H_{\mathring{p}}\mathring{\mathrm{E}} = \\ = H_{\mathring{p}}\boldsymbol{\mu_p}\boldsymbol{1_n}^\top + BU + \mathrm{E},\ \mathrm{E} \sim \mathcal{MN}(0, I_p, D). \end{aligned} \tag{32}$$

From this time forward, we shall use this transformed model instead of Equation 10:

$$Y = H_p\boldsymbol{\mu_p}\boldsymbol{1_n}^\top + BU + \mathrm{E},\ U \sim \mathcal{MN}(\boldsymbol{\mu_m}\boldsymbol{1_n}^\top, \Sigma, G),\ \mathrm{E} \sim \mathcal{MN}(0, I_p, D). \tag{33}$$

## 4.3. Dealing with $\boldsymbol{\mu_p}$ and $\boldsymbol{\mu_m}$

By transforming the Equation 34 from the right by $H_n^\top$, we get

$$YH_n^\top = \tilde{Y} = \cancel{H_{\mathring{p}}\boldsymbol{\mu_p}\boldsymbol{1_n}^\top H_n} + BUH_n + EH_n = B\tilde{U} + \tilde{\mathrm{E}},\ \tilde{U} \sim \mathcal{MN}(0, \Sigma, H_nGH_n^\top),\ \mathrm{E} \sim \mathcal{MN}(0, I_p, H_nDH_n^\top). \tag{34}$$

## 4.4. Estimating $D$ with Restricted Maximum Likelihood (REML)

REML idea is to transform the Equation 34 to get rid of the $B\tilde{U}$ term, i.e. we need to find such matrix $P$ that $P\tilde{Y} = PB\tilde{U} + P\tilde{\mathrm{E}} = P\tilde{\mathrm{E}}$. Surely, there are plenty of matrices such that $PB = 0$, but we seek a transformation of a maximal rank. To this end, consider a singular value decomposition (SVD) of the $B$ matrix of rank $r$:

$$B = \underbrace{Q}_{p\times p}\ \underbrace{S}_{p\times m}\ \underbrace{V}_{m\times m} = \tag{35}$$

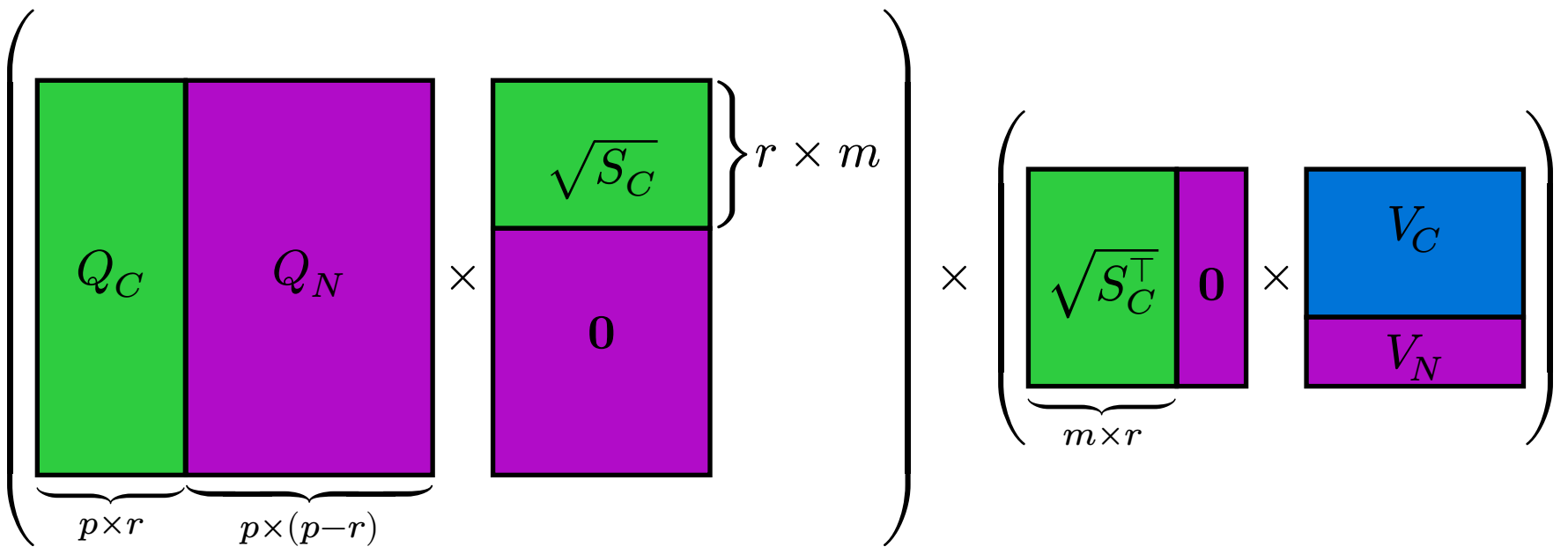

$$= \left(Q_C\sqrt{S_C} + \cancel{Q_N\mathbf{0}}\right)\left(\sqrt{S_C^\top}V_C + \cancel{\mathbf{0}V_N}\right) = Q_CS_CV_C.$$

As $Q_c \perp Q_N$, we can pick $P = Q_N^\top$:

$$Q_N^\top\tilde{Y} = \hat{Y} = Q_NB\tilde{U} + Q_N\tilde{\mathrm{E}} = Q_N\tilde{\mathrm{E}} = \hat{\mathrm{E}},\ \hat{\mathrm{E}} \sim \mathcal{MN}\left(0, \overbrace{I_{p-r}}^{=Q_N^\top Q_N}, H_nDH_n^\top\right). \tag{36}$$

The log-likelihood is:

$$l\left(\hat{Y}|D\right) = \mathrm{tr}\left\{\hat{Y}\left(H_n D H_n^\top\right)^{-1}\hat{Y}^\top\right\} + \frac{p-r}{2}\ln\det|H_n D H_n^\top| \tag{37}$$

Unless there is a single group (hence $D \propto I_n$), $D$ can't be estimated analytically from Equation 37. Instead, we use gradient-descent based numerical scheme.

#### 4.4.1. Efficient computation of $\hat{Y}$

The process of computing the full SVD decomposition of $B$ can be computationally intensive. In practice, it is only possible to employ the reduced SVD $B = Q_C S_C V_C$; however, the orthogonal complement $Q_N$ is still required in order to transform the data matrix $Y$. The primary issue is that $Q_N$ possesses a shape $p \times (p-r)$, which necessitates $O(p^2)$ storage. This approach becomes impractical when the number of genes, $p$, is substantial. Therefore, we propose a method to efficiently estimate $\hat{Y} = Q_N^T \tilde{Y}$ without forming the $Q_N$ matrix explicitly, based on Householder transformations.

A Householder transformation is:

$$\mathcal{H} = I - 2\frac{\boldsymbol{u}\boldsymbol{u}^\top}{\boldsymbol{u}^\top\boldsymbol{u}}, \tag{38}$$

where $\boldsymbol{u}$ is non-zero vector, and $\mathcal{H}$ is an orthogonal matrix . We construct $Q = [Q_C, Q_N]$ as:

$$Q = \mathcal{H}_1\mathcal{H}_2 \ldots \mathcal{H}_r, \tag{39}$$

so $Q$'s first $r$ columns are $Q_C$, i.e., $Q\boldsymbol{e}_j = \boldsymbol{q}_{C,j}$ for $j = 1, \ldots, r$, where $\boldsymbol{e}_j$ is the $j$-th standard basis vector and $\boldsymbol{q}_{C,j}$ is the $j$-th column of $Q_C$. The remaining columns form $Q_N$. It is easy to compute matrix-matrix products if one of the matrices is $\mathcal{I}$ given its convenient representation in terms of a diagonal identity matrix and an outer vector product.

Let's assume now that we have the necessary $\{\mathcal{H}_i\}_{\{i=1\}}^r$ at a hand. Then, if we transformed $\tilde{Y}$ by $Q^\top$,

$$Q^\top\tilde{Y} = \left(Q_C \;\; Q_N\right)^\top\tilde{Y} = \begin{pmatrix} Q_C^\top\tilde{Y} \\ Q_N^\top\tilde{Y} \end{pmatrix}, \tag{40}$$

we could have just extracted the last $p-r$ rows of this quantity to obtain $Q_N^\top\tilde{Y}$. Using the Householder representation of $Q$,

$$Q^\top\tilde{Y} = \mathcal{H}_r\left(\mathcal{H}_{r-1}\left(\ldots\left(\mathcal{H}_1\tilde{Y}\right)\ldots\right)\right). \tag{41}$$

Still, we need to obtain $\{\mathcal{H}_i\}_{\{i=1\}}^r$ first. Each $\mathcal{H}_j$ acts on dimensions $j$ to $p$:

$$\mathcal{H}_j = I - 2\frac{\boldsymbol{u_j}\boldsymbol{u_j}^\top}{\boldsymbol{u_j}^\top\boldsymbol{u_j}} = \begin{pmatrix} I_{j-1} & 0 \\ 0 & I_{p-j+1} - 2\frac{\hat{\boldsymbol{u}}_j\hat{\boldsymbol{u}}_j^T}{\hat{\boldsymbol{u}}_j^T\hat{\boldsymbol{u}}_j} \end{pmatrix}, \tag{42}$$

where $\hat{\boldsymbol{u}}_j$ is a $(p-j+1)$-vector, and $\boldsymbol{u}_j = \left[0_{j-1}, \hat{\boldsymbol{u}}_j\right]$. To compute $\boldsymbol{u}_j$, we ensure that $\mathcal{H}_j$ maps $\boldsymbol{x}_j = \mathcal{H}_{j-1}\ldots\mathcal{H}_1\boldsymbol{e}_j$ to $\boldsymbol{q}_{C,j}$. Let's define

$$\boldsymbol{u}_j = \boldsymbol{x}_j - s_j\|\boldsymbol{x}_j\|\boldsymbol{q}_{C,j}, \tag{43}$$

where $\|\boldsymbol{x}_j\| = \sqrt{\boldsymbol{x}_j^T\boldsymbol{x}_j}$ and $s_j = \mathrm{sign}\left(\boldsymbol{x}_{j[j]}\right)\ \mathrm{sign}\left(\boldsymbol{q}_{C,j}[j]\right)$ helps avoiding cancellation. Then:

$$\boldsymbol{u}_{j[j:p]} = \boldsymbol{x}_{j[j:p]} - s_j\boldsymbol{q}_{C,j}[j:p], \quad \boldsymbol{u}_{j[1:j-1]} = 0. \tag{44}$$

Next, we normalize $\hat{\boldsymbol{u}}_j$ so that $\hat{\boldsymbol{u}}_{j[0]} = 1$, scaling $\hat{\boldsymbol{u}}_{j[1:p-j+1]}$ by dividing it by $\hat{\boldsymbol{u}}_{j[0]}$. This process repeats, updating $\boldsymbol{x}_{j+1}$ via prior $\mathcal{H}_j$.

This approach requires just $O(pr)$ storage and has a total $O(prn)$ complexity.

#### 4.4.2. Log-likelihood computation

Computing Equation 37 requires inverting the matrix $(n-1) \times (n-1)$ matrix and computing its determinant. Those steps have a complexity of $O((n-1)^3)$ each, but, being thankful to Property 4.1.2.3 and Property 4.1.2.2, the overall complexity can be reduced to linear in $n$. The $\ln\det|H_n D H_n^\top|$ is trivially computed with Property 4.1.2.3:

$$\ln\det|H_n D H_n^\top| = \ln|D| + \ln\left(\sum_{i=1}^n \frac{1}{d_i}\right) - \ln(n), \tag{45}$$

Table 1: Reminder for the notation used throughout the paper.

| Symbol | Description |
|---|---|
| $\mathring{p}$ | Number of promoters/genes |
| $n$ | Numer of samples across all groups |
| $g$ | Number of groups |
| $m$ | Number of motifs |
| $\mathring{Y}$ | An $p \times n$ matrix of $\log$ expression values |
| $Y$ | $Y = H_{\mathring{p}}\mathring{Y}$ |
| $\tilde{Y}$ | $\tilde{Y} = YH_n^\top$ |
| $\hat{Y}$ | $\hat{Y} = Q_N\tilde{Y}$ |
| $\check{Y}$ | $\check{Y} = \hat{Y}L^{-T}$, where $L$ is a Cholesky factor of $H_nDH_n^\top = LL^\top$ |
| $\mathring{B}$ | A loading matrix of motifs onto genes/ promoters |
| $B$ | $B = H_{\mathring{p}}\mathring{B}$ |
| $\mathring{\mathrm{E}}$ | A random matrix of errors/noise |
| $\mathrm{E}$ | $\mathrm{E} = H_{\mathring{p}}\mathring{\mathrm{E}}$ |
| $\hat{\mathrm{E}}$ | $\hat{\mathrm{E}} = \mathrm{E}H_n^\top$ |
| $U$ | A random matrix of motif activities |
| $\hat{U}$ | $\hat{U} = UH_n^\top$ |

| Symbol | Description |
|---|---|
| $p$ | $p = \mathring{p} - 1$ |
| $r$ | Rank of $B$ matrix |
| $\boldsymbol{\mu_p}$ | A vector of promoter-wise means |
| $\boldsymbol{\mu_s}$ | A vector of sample-wise means |
| $\boldsymbol{\mu_m}$ | A vector of motif activities means |
| $\boldsymbol{\mu_s}$ | A vector of sample-wise means |
| $H_j$ | The semi-orthogonal complement matrix of size $j - 1 \times j$ |
| $Q_N$ | A semi-orthogonal matrix such that $Q_NB = 0$. Constructed from an SVD decomposition of $B = [Q_C, Q_N]D_BV_B$ |
| $D$ | Diagonal matrix of $g$ unique variance parameters of size $n$, variance of noise $\mathrm{E}$ |
| $G$ | Diagonal matrix of $g$ unique variance parameters of size $n$, effectively contains group-wise multipliers of random motif effects per group |
| $\Sigma$ | Diagonal matrix of motif variances of size $m$ |

whereas for the inverse matrix we have to remember that $\hat{Y} = Q_N^\top \tilde{Y} H_n^\top$. Lets re-denote as $Q_N^\top \tilde{Y}$ as $\check{Y}$ here. At this point, we used accent notation a lot and it might become confusing – in that case, take a look at Table 1 Then, the trace term attains the form:

$$\operatorname{tr}\Big\{\check{Y}H_n^\top\big(H_nDH_n^\top\big)^{-1}H_n\check{Y}^\top\Big\} = \overbrace{\frac{1}{n}\frac{\det|D|}{\det|H_nDH_n^\top|}}^{\xi}\operatorname{tr}\Big\{\check{Y}(K - \boldsymbol{d}^{-1}\boldsymbol{d}^{-T})\check{Y}^\top\Big\} = \\ = \xi\operatorname{tr}\Big\{\check{Y}K\check{Y}^\top - \check{Y}\boldsymbol{d}^{-1}\boldsymbol{d}^{-T}\check{Y}^\top\Big\}. \tag{46}$$

### 4.4.3. Gradient calculation

$$\frac{\partial l\big(\check{Y}|D\big)}{\partial\sigma_k} = \frac{p-r}{2}\operatorname{tr}\Big\{\big(H_nDH_n^\top\big)^{-1}H_n\frac{\partial D}{\partial\sigma_k}H_n^\top\Big\} - \operatorname{tr}\Big\{\check{Y}\Big[H_n^\top\big(H_nDH_n^\top\big)^{-1}H_n\Big]\frac{\partial D}{\partial\sigma_k}\Big[H_n^\top\big(H_nDH_n^\top\big)^{-1}H_n^\top\Big]\check{Y}^\top\Big\} = \\ = \frac{p-r}{2}\xi\operatorname{tr}\Big\{(K - \boldsymbol{d}^{-1}\boldsymbol{d}^{-T})\frac{\partial D}{\partial\sigma_k}\Big\} - \xi^2\operatorname{tr}\left\{\underbrace{\check{Y}(K - \boldsymbol{d}^{-1}\boldsymbol{d}^{-T})}_{Z}\frac{\partial D}{\partial\sigma_k}(K - \boldsymbol{d}^{-1}\boldsymbol{d}^{-T})\check{Y}^\top\right\} = \\ = \frac{p-r}{2}\xi\operatorname{tr}\Big\{(K - \boldsymbol{d}^{-1}\boldsymbol{d}^{-T})\frac{\partial D}{\partial\sigma_k}\Big\} - \xi^2\operatorname{tr}\Big\{Z^\top Z\frac{\partial D}{\partial\sigma_k}\Big\}, \tag{47}$$

where $\frac{\partial D}{\partial\sigma_k} = \mathcal{I}_{\mathcal{G}(\not{k})}$ is an indicator matrix of size $g \times n$ for the $k$-th group. Then, in a vectorized form:

$$\frac{\partial l\big(\check{Y}|D\big)}{\partial\boldsymbol{\sigma}} = \mathcal{I}_{\mathcal{G}(\not{k})}\Big(\frac{p-r}{2}\xi\ \mathtt{diag}[K - \boldsymbol{d}^{-1}\boldsymbol{d}^{-T}] - \xi^2\ \mathtt{diag}[Z^\top Z]\Big). \tag{48}$$

## 4.5. Estimating $\boldsymbol{\mu_p}$

Let's continue to use $\check{Y}$ as defined in the previous section, i.e. $\mathring{Y}$ left-transformed by $H_p$ and then by $Q_N^\top$:

$$\check{Y} = \underbrace{Q_N^\top H_p}_{A}\boldsymbol{\mu_p}\mathbf{1_n}^\top + \mathrm{E}, \mathrm{E} \sim \mathcal{MN}(0, I_r, D). \tag{49}$$

Likewise, we use ML to estimate $\mu_p$:

$$l\left(\breve{Y}|\boldsymbol{\mu_p}\right) = \mathrm{tr}\left\{\left(\breve{Y} - A\boldsymbol{\mu_p}\mathbf{1_n}^\top\right)D^{-1}\left(\breve{Y} - A\boldsymbol{\mu_p}\mathbf{1_n}^\top\right)^\top\right\} = \mathrm{tr}\left\{\breve{Y}D^{-1}\breve{Y}^\top + A\boldsymbol{\mu_p}\mathbf{1_n}^\top D^{-1}\mathbf{1_n}\boldsymbol{\mu_p}^\top A^\top - \right.$$
$$\left. -2A\boldsymbol{\mu_p}\mathbf{1_n}^\top D^{-1}\breve{Y}^\top\right\} = \mathrm{tr}\left\{\breve{Y}D^{-1}\breve{Y}^\top + \omega A\boldsymbol{\mu_p}\boldsymbol{\mu_p}^\top A^\top - 2A\boldsymbol{\mu_p}\mathbf{1_n}^\top D^{-1}\breve{Y}^\top\right\}, \tag{50}$$

where $\omega = \mathbf{1_n}^\top D^{-1}\mathbf{1_n} = \sum_{i=1}^k \frac{1}{\sigma_i}$

$$\frac{\partial l\left(\breve{Y}\ |\boldsymbol{\mu_p}\right)}{\partial \boldsymbol{\mu_p}} = \omega A^\top A\boldsymbol{\mu_p} - A^\top \breve{Y}D^{-1}\mathbf{1_n} = 0 \leftrightarrow \omega A^\top A\boldsymbol{\mu_p} = A^\top \breve{Y}\boldsymbol{d}^{-1} \leftrightarrow A^\top A\boldsymbol{\mu_p} = \frac{1}{\omega}A^\top \breve{Y}\boldsymbol{d}^{-1} = \frac{1}{\omega}A^\top A\mathring{Y}\boldsymbol{d}^{-1}. \tag{51}$$

One possible solution for the equation is $\boldsymbol{\mu_p} = \frac{1}{\mathbf{1_n}^\top D^{-1}\mathbf{1_n}}\mathring{Y}D^{-1}\mathbf{1_n}$, i.e. the mean across promoters is just a reciprocal variance weighted average. However, see that the number of possible solutions is, in fact, infinite, as $A^\top A$ is not of a full rank. To avoid large values in the estimates of $\boldsymbol{\mu_p}$, we instead look for a minimum norm solution that is achievable using Moore-Penrose inverse:

$$\boldsymbol{\mu_p} = \frac{1}{\mathbf{1_n}^\top D^{-1}\mathbf{1_n}}H_{\mathring{p}}^\top Q_N Q_N^\top Y D^{-1}\mathbf{1_n} = \frac{1}{\mathbf{1_n}^\top D^{-1}\mathbf{1_n}}H_{\mathring{p}}^\top (I_{\mathring{p}} - Q_C Q_C^\top)YD^{-1}\mathbf{1_n}, \tag{52}$$

which can be computed efficiently by avoiding matrix-matrix products and only evaluating matrix-vector products. The minimum-norm assumption is a necessary prior that makes estimating $\mu_m$ possible later on.

## 4.6. Estimating $\Sigma$ and $G$

We continue working with the doubly “centered” $Y$, but this time we do not apply the $Q_N^\top$ transformation to keep $U$ term intact:

$$\tilde{Y} = B\widehat{UG} + \hat{\mathrm{E}},\ \hat{U} \sim \mathcal{MN}(0, \Sigma, H_n G H_n^\top),\ \hat{\mathrm{E}} \sim \mathcal{MN}(0, I_p, H_n D H_n^\top). \tag{53}$$

What’s the distribution of circle $\mathring{Y}$? Due to Equation 1 it is certainly normal:

$$\mathsf{vec}\left[\tilde{Y}\right] \sim \mathcal{N}\left(0, (H_n G H_n^\top) \otimes (B\Sigma B^\top) + (H_n D H_n^\top) \otimes I_p\right). \tag{54}$$

There is nothing wrong with this form, but in practice inverting its covariance matrix of size $np \times np$ is not feasible. Instead, first, we obtain the Cholesky decomposition of $H_n D H_n^\top = LL^\top$ (alternatively, any kind of $LL^\top$ decomposition can be used here, as long as $L$ is invertible). Then, we transform $Y$ with $L^{-T}$ from the right:

$$\tilde{Y}L^{-T} = \breve{Y} = B\tilde{U} + \breve{\mathrm{E}},\ \tilde{U} \sim \mathcal{MN}(0, \Sigma, L^{-1}H_n G H_n^\top L^{-T}), \breve{\mathrm{E}} \sim \mathcal{MN}(0, I_p, I_{n-1}). \tag{55}$$

Then,

$$\mathsf{vec}\left[\breve{Y}\right] \sim \mathcal{N}\left(0, (L^{-1}H_n G H_n^\top L^{-T}) \otimes (B\Sigma B^\top) + I_{p(n-1)}\right). \tag{56}$$

We shall reap benefits of this representation shortly.

### 4.6.1. Loglikelihood computation

Naively, loglikelihood of Equation 55 is

$$l\left(\breve{Y}|\Sigma, G\right) = \mathsf{vec}\left[\breve{Y}\right]^\top \left((L^{-1}H_n G H_n^\top L^{-T}) \otimes (B\Sigma B^\top) + I_{p(n-1)}\right)^{-1} \mathsf{vec}\left[\breve{Y}\right] +$$
$$+ \ln\det|(L^{-1}H_n G H_n^\top L^{-T}) \otimes (B\Sigma B^\top) + I_{p(n-1)}|. \tag{57}$$

First, let’s detnote $L^{-1}H_n\sqrt{G}$ as A and $B\sqrt{\Sigma}$ as $C$. The covariance matrix is then $(AA^\top) \otimes (CC^\top) + I_{(p-1)(n-1)}$. Furthermore, using the mixed-product property of the kronecker product, $(\mathrm{A}\mathrm{A}^\top) \otimes (CC^\top) = \underbrace{(\mathrm{A} \otimes C)}_{V}(A^\top \otimes C^\top) = VV^\top$. Now it is easy to recognize that the covariance matrix has form that makes application of Equation 3 and Equation 4 possible:

$$\left(VV^\top + I_{p(n-1)}\right)^{-1} = I_{p(n-1)} - V\underbrace{(I_{mn} + V^\top V)}_{S}{}^{-1}V^\top = I_{p(n-1)} - VS^{-1}V^\top, \tag{58}$$

where

$$S = V^\top V + I_{nm} = (\mathrm{A}^\top \otimes C^\top)(\mathrm{A} \otimes C) + I_{nm} = (A^\top A) \otimes (C^\top C) + I_{nm}. \tag{59}$$

Furthermore, now we can take advantage of Equation 8 as an eigenvalue decomposition of $C^\top C = \sqrt{\Sigma} B^\top B \sqrt{\Sigma} = Q_B D_B Q_B^\top$ is feasible ($C^\top C$ matrix has dimensions $m \times m$, whereas $CC^\top$ is $p \times (p-1)$). As for $A^\top A$, see that

$$A^\top A = \sqrt{G} H_n^\top L^{-T} L^{-1} H_n \sqrt{G} = \sqrt{G} H_n^\top (H_n D H_n)^{-1} H_n \sqrt{G}. \tag{60}$$

Here, we can take advantage of Property 4.1.2.2:

$$A^\top A = \frac{1}{n} \frac{\det|D|}{\det|H_n D H_n^\top|} \sqrt{G} (K - \boldsymbol{d}^{-1} \boldsymbol{d}^{-T}) \sqrt{G}. \tag{61}$$

This representation allows us to compute matrix-vector products in $O(n)$, as $G$ and $K$ are diagonal matrices and $\boldsymbol{d}^{-1}$ is a vector, and greatly fascilates numerical speed of numerical algorithms for eigenvalue decompositions.

After we get eigenvalue decompositions for $A^\top A$ and $C^\top C$, we can easily compute the inverse of $S$:

$$S^{-1} = (Q_A \otimes Q_B)(D_A \otimes D_B + I_{nm})^{-1} (Q_A^\top \otimes Q_B^\top). \tag{62}$$

Then, the determinant is trivially computed. As for the quadratic form term,

$$\begin{aligned}
&\operatorname{vec}\left[\check{Y}\right]^\top V (Q_A \otimes Q_B)(D_A \otimes D_B + I_{nm})^{-1} (Q_A^\top \otimes Q_B^\top) V^\top \operatorname{vec}\left[\check{Y}\right] = \\
&= \operatorname{vec}\left[\sqrt{\Sigma} B^\top \check{Y} L^{-1} H_n \sqrt{G}\right]^\top (Q_A \otimes Q_B)(D_A \otimes D_B + I_{nm})^{-1} (Q_A^\top \otimes Q_B^\top) \operatorname{vec}\left[\sqrt{\Sigma} B^\top \check{Y} L^{-1} H_n \sqrt{G}\right] = \\
&= \operatorname{vec}\left[Q_B^\top \sqrt{\Sigma} \underbrace{B^\top \check{Y} L^{-1} H_n}_{Z} \sqrt{G} Q_A\right]^\top (D_A \otimes D_B + I_{nm})^{-1} \operatorname{vec}\left[Q_B^\top \sqrt{\Sigma} B^\top \check{Y} L^{-1} H_n \sqrt{G} Q_A\right] = \\
&= \operatorname{vec}\left[Q_B^\top \sqrt{\Sigma} Z \sqrt{G} Q_A\right]^\top (D_A \otimes D_B + I_{nm})^{-1} \operatorname{vec}\left[Q_B^\top \sqrt{\Sigma} Z \sqrt{G} Q_A\right],
\end{aligned} \tag{63}$$

where $Z = B^\top \check{Y} L^{-1} H_n = B^\top Y H_n^\top L^{-T} L^{-1} H_n = B^\top Y H_n^\top (H_n D H_n^\top)^{-1} H_n$ has to be computed once.

This way, the computation of the loglikelihood requires no operations cubic in $n$ or $p$.

#### 4.6.2. Gradient computation

Let's rewrite Equation 57 in terms of $A = L^{-1} H_n \sqrt{G}$ and $C = B\sqrt{\Sigma}$ matrices for brevity:

$$l\left(\check{Y}, \Sigma, G\right) = \operatorname{vec}\left[\check{Y}\right]^\top \underbrace{\left((AA^\top) \otimes (CC^\top) + I_{p(n-1)}\right)}_{\hat{S}}^{-1} \operatorname{vec}\left[\check{Y}\right] + \ln \det|(A^\top A) \otimes (C^\top C) + I_{nm}|. \tag{64}$$

If parameter of interest $\theta$ resides in $AA^\top$, then $\frac{\partial S}{\partial \theta} = \frac{\partial (AA^\top)}{\partial \theta} \otimes (CC^\top)$, otherwise $\frac{\partial S}{\partial \theta} = (AA^\top) \otimes \frac{\partial (CC^\top)}{\partial \theta}$.

##### 4.6.2.1. Derivative w.r.t. parameter $\tau_i$ in $\Sigma$

###### 4.6.2.1.1. Derivative of the quadratic form

$$\begin{aligned}
-\frac{\partial \operatorname{vec}\left[\check{Y}\right]^\top \hat{S}^{-1} \operatorname{vec}\left[\check{Y}\right]}{\partial \tau_i} &= \operatorname{vec}\left[\check{Y}\right]^\top \hat{S}^{-1} (AA^\top) \otimes \left(B \frac{\partial \Sigma}{\partial \tau_i} B^\top\right) \hat{S}^{-1} \operatorname{vec}\left[\check{Y}\right] = \\
&= \operatorname{vec}\left[\check{Y}\right] \hat{S}^{-1} (A \otimes (BJ_i))(A^\top \otimes (J_i B^\top)) \hat{S}^{-1} \operatorname{vec}\left[\check{Y}\right] = \\
&= \operatorname{vec}\left[\check{Y}\right] \hat{S}^{-1} (A \otimes (BJ_i)) \left(\hat{S}^{-1} (A \otimes (BJ_i))\right)^\top \operatorname{vec}\left[\check{Y}\right].
\end{aligned} \tag{65}$$

Let's rewrite $\hat{S}^{-1}(A \otimes (BJ_i))$ as

$$\hat{S}^{-1}(A \otimes (BJ_i)) = \hat{S}^{-1}\left(A \otimes \left(B\sqrt{\Sigma}\sqrt{\Sigma}^{-1} J_i\right)\right) = \hat{S}^{-1}\left(A \otimes \left(B\sqrt{\Sigma}\right)\right)\left(I_n \otimes \left(\sqrt{\Sigma}^{-1} J_i\right)\right). \tag{66}$$

Then, we can use Equation 2:

$$\ldots = \left(A \otimes \left(B\sqrt{\Sigma}\right)\right) \underbrace{\left((A^\top A) \otimes (C^\top C) + I_{nm}\right)}_{S^{-1}} \left(I_n \otimes \left(\sqrt{\Sigma}^{-1} J_i\right)\right) = \left(A \otimes \left(B\sqrt{\Sigma}\right)\right) S^{-1} \left(I_n \otimes \left(\sqrt{\Sigma}^{-1} J_i\right)\right). \tag{67}$$

Plugging it into the quadratic product formula:

$$\ldots = \operatorname{vec}\left[\check{Y}\right]^\top \left(A \otimes \left(B\sqrt{\Sigma}\right)\right) S^{-1} \left(I_n \otimes \left(\sqrt{\Sigma}^{-1} J_i \sqrt{\Sigma}^{-1}\right)\right) S^{-1} \left(A^\top \otimes \left(B\sqrt{\Sigma}\right)^\top\right) \operatorname{vec}\left[\check{Y}\right] = \tag{68}$$

$$= \mathtt{vec}\left[\sqrt{\Sigma}B^\top \check{Y}A\right]^\top S^{-1}\left(I_n \otimes \left(\sqrt{\Sigma}^{-1}J_i\sqrt{\Sigma}^{-1}\right)\right)S^{-1}\ \mathtt{vec}\left[\sqrt{\Sigma}B^\top \check{Y}A\right]. \tag{68}$$

Remembering Equation 62:

$$\begin{gathered}
\mathtt{vec}\left[\sqrt{\Sigma}B^\top \check{Y}A\right]^\top S^{-1}\left(I_n \otimes \left(\sqrt{\Sigma}^{-1}J_i\sqrt{\Sigma}^{-1}\right)\right)S^{-1}\ \mathtt{vec}\left[\sqrt{\Sigma}B^\top \check{Y}A\right] = \\
\mathtt{vec}\left[\sqrt{\Sigma}B^\top \check{Y}A\right]^\top (Q_A \otimes Q_B)(D_A \otimes D_B + I_{nm})^{-1} \times \\
\times\ (Q_A^\top \otimes Q_B^\top)\left(I_n \otimes \left(\sqrt{\Sigma}^{-1}J_i\sqrt{\Sigma}^{-1}\right)\right)(Q_A \otimes Q_B)(D_A \otimes D_B + I_{nm})^{-1}(Q_A^\top \otimes Q_B^\top)\ \mathtt{vec}\left[\sqrt{\Sigma}B^\top \check{Y}A\right] = \\
\mathtt{vec}\left[Q_B^\top\sqrt{\Sigma}B^\top \check{Y}AQ_A\right]^\top (D_A \otimes D_B + I_{nm})^{-1}\left(I_n \otimes \left(Q_B^\top\left(\sqrt{\Sigma}^{-1}J_iJ_i\sqrt{\Sigma}^{-1}\right)Q_B\right)\right) \times \\
\times\ (D_A \otimes D_B + I_{nm})^{-1}\ \mathtt{vec}\left[Q_B^\top\sqrt{\Sigma}B^\top \check{Y}AQ_A\right] = \\
\frac{1}{\tau_i}\mathtt{vec}\left[Q_B^\top\sqrt{\Sigma}B^\top \check{Y}AQ_A\right]^\top (D_A \otimes D_B + I_{nm})^{-1}(I_n \otimes (Q_B^\top J_i Q_B))(D_A \otimes D_B + I_{nm})^{-1}\ \mathtt{vec}\left[Q_B^\top\sqrt{\Sigma}B^\top \check{Y}AQ_A\right].
\end{gathered} \tag{69}$$

Let's denote $M = \mathtt{mat}_m^n\left[\mathtt{diag}\left[(D_A \otimes D_B + I_{nm})^{-1}\right]\right]$ (i.e. we first extracted the non-zero diagonal of the diagonal matrix $(D_A \otimes D_B + I_{nm})^{-1}$ and then refolded it into a matrix of size $m \times n$), then:

$$\begin{aligned}
\ldots &= \frac{1}{\tau_i}\mathtt{vec}\left[\left(\left(Q_B^\top\sqrt{\Sigma}B^\top \check{Y}AQ_A\right) \odot M\right)\right]^\top (I_n \otimes (Q_B^\top J_i Q_B))\ \mathtt{vec}\left[\left(\left(Q_B^\top\sqrt{\Sigma}B^\top \check{Y}AQ_A\right) \odot M\right)\right] = \\
&= \frac{1}{\tau_i}\mathtt{vec}\left[\left(\left(Q_B^\top\sqrt{\Sigma}B^\top \check{Y}AQ_A\right) \odot M\right)\right]^\top (I_n \otimes (Q_B^\top J_i))(I_n \otimes (J_i Q_B))\ \mathtt{vec}\left[\left(\left(Q_B^\top\sqrt{\Sigma}B^\top \check{Y}AQ_A\right) \odot M\right)\right] = \\
&= \frac{1}{\tau_i}\mathtt{vec}\left[J_iQ_B\left(\left(Q_B^\top\sqrt{\Sigma}B^\top \check{Y}AQ_A\right) \odot M\right)\right]^\top\ \mathtt{vec}\left[J_iQ_B\left(\left(Q_B^\top\sqrt{\Sigma}B^\top \check{Y}AQ_A\right) \odot M\right)\right] = \\
&= \frac{1}{\tau_i}\,\mathrm{tr}\left\{\left(\left(Q_B^\top\sqrt{\Sigma}B^\top \check{Y}AQ_A\right) \odot M\right)^\top Q_B^\top J_i Q_B\left(\left(Q_B^\top\sqrt{\Sigma}B^\top \check{Y}AQ_A\right) \odot M\right)\right\} = \\
&= \frac{1}{\tau_i}\,\mathrm{tr}\left\{Q_B\left(\left(Q_B^\top\sqrt{\Sigma}Z\sqrt{G}Q_A\right) \odot M\right)\left(\left(Q_B^\top\sqrt{\Sigma}Z\sqrt{G}Q_A\right) \odot M\right)^\top Q_B^\top J_i\right\} \\
&= \frac{1}{\tau_i}\,\mathrm{tr}\left\{\underbrace{Q_B\left(\left(Q_B^\top\sqrt{\Sigma}Z\sqrt{G}Q_A\right) \odot M\right)}_{\Lambda}\left(\left(Q_B^\top\sqrt{\Sigma}Z\sqrt{G}Q_A\right) \odot M\right)^\top Q_B^\top J_i\right\} = \frac{1}{\tau_i}\,\mathrm{tr}\{\Lambda\Lambda^\top J_i\},
\end{aligned} \tag{70}$$

or in a vector form:

$$\frac{\partial \mathtt{vec}\left[\check{Y}\right]^\top \hat{S}^{-1}\ \mathtt{vec}\left[\check{Y}\right]}{\partial \boldsymbol{\tau}} = -\boldsymbol{\tau}^{-1} \odot \mathtt{diag}[\Lambda\Lambda^\top]. \tag{71}$$

#### 4.6.2.1.2. Derivative of the determinant term

$$\begin{aligned}
\frac{\partial \ln\det \hat{S}}{\partial \tau_i} &= \frac{\partial \ln\det S}{\partial \tau_i} = \mathrm{tr}\left\{S^{-1}\frac{\partial S}{\partial \tau_i}\right\} = \mathrm{tr}\left\{S^{-1}(A^\top A) \otimes \left(\frac{\partial\sqrt{\Sigma}B^\top B\sqrt{\Sigma}}{\partial \tau_i}\right)\right\} = \\
&= \frac{1}{2\sqrt{\tau_i}}\,\mathrm{tr}\left\{S^{-1}(A^\top A) \otimes \left(J_iB^\top B\sqrt{\Sigma} + \sqrt{\Sigma}B^\top BJ_i\right)\right\} = \frac{1}{\sqrt{\tau_i}}\,\mathrm{tr}\left\{S^{-1}(A^\top A) \otimes \left(\sqrt{\Sigma}B^\top BJ_i\right)\right\} = \\
&= \frac{1}{\sqrt{\tau_i}}\,\mathrm{tr}\left\{(Q_A \otimes Q_B)(D_A \otimes D_B + I_{nm})^{-1}(Q_A^\top \otimes Q_B^\top)\left((Q_AD_AQ_A^\top) \otimes \left(\sqrt{\Sigma}B^\top BJ_i\right)\right)\right\} = \\
&= \frac{1}{\sqrt{\tau_i}}\,\mathrm{tr}\left\{(D_A \otimes D_B + I_{nm})^{-1}\left(D_A \otimes \left(Q_B^\top\sqrt{\Sigma}B^\top BJ_iQ_B\right)\right)\right\}.
\end{aligned} \tag{72}$$

If we denote $\mathtt{diag}\left[(D_A \otimes D_B + I_{nm})^{-1}\right] = \boldsymbol{s}$, $\mathtt{diag}[D_A] = \boldsymbol{d_A}$ and $\mathtt{diag}\left[Q_B^\top\sqrt{\Sigma}B^\top BJ_iQ_B\right] = \boldsymbol{\zeta_j}$, then

$$\frac{\partial \ln\det \hat{S}}{\partial \tau_i} = \frac{1}{\sqrt{\tau_j}}\mathbf{1}_{nm}^\top(\boldsymbol{s} \odot (\boldsymbol{d_A} \otimes \boldsymbol{\zeta_j})) = \frac{1}{\sqrt{\tau_j}}\boldsymbol{s}^\top(\boldsymbol{d_A} \otimes \boldsymbol{\zeta_j}). \tag{73}$$

To generalize it in terms of derivative w.r.t. vector $\boldsymbol{\tau}$, we care re-arrange $\zeta_j$ as $\zeta_j = \left[\underbrace{\left(Q_B^\top\sqrt{\Sigma}B^\top B\right) \odot Q_B^\top}_{\mathrm{Z}}\right]_{[,j]} = \mathrm{Z}_{.,j}$. Then:

$$\frac{\partial \ln\det\hat{S}}{\partial \boldsymbol{\tau}} = \frac{1}{\sqrt{\boldsymbol{\tau}}} \odot \boldsymbol{s}^\top(\boldsymbol{d_A} \otimes \mathrm{Z}). \tag{74}$$

#### 4.6.2.2. Derivative w.r.t. parameter $\nu$ in $G$

##### 4.6.2.2.1. Derivative of the quadratic form

$$\begin{aligned}
-\frac{\partial \mathsf{vec}\left[\check{Y}\right]^\top \hat{S}^{-1}\ \mathsf{vec}\left[\check{Y}\right]}{\partial \nu_i} &= \mathsf{vec}\left[\check{Y}\right]^\top \hat{S}^{-1}\left(L^{-1}H_n J_{\mathcal{G}(i)} H_n^\top L^{-T}\right) \otimes (CC^\top)\hat{S}^{-1}\ \mathsf{vec}\left[\check{Y}\right] = \\
= \mathsf{vec}\left[\check{Y}\right]^\top \hat{S}^{-1}&\left(\left(L^{-1}H_n\sqrt{G}\right)\otimes C\right)\left(\sqrt{G}^{-1} J_{\mathcal{G}(i)}\sqrt{G}^{-1}\otimes I_m\right)\left(\left(\sqrt{G}H_n L^{-1}\right)\otimes C^\top\right)\hat{S}^{-1}\ \mathsf{vec}\left[\check{Y}\right] = \\
= \frac{1}{\nu_i}\mathsf{vec}\left[\check{Y}\right]^\top &\left(\left(L^{-1}H_n\sqrt{G}\right)\otimes C\right)S^{-1}\left(J_{\mathcal{G}(i)}\otimes I_m\right)S^{-1}\left(\left(\sqrt{G}H_n^\top L^{-T}\right)\otimes C^\top\right)\ \mathsf{vec}\left[\check{Y}\right] = \\
= \frac{1}{\nu_i}\mathsf{vec}&\left[C^\top\check{Y}L^{-1}H_n\sqrt{G}\right]^\top S^{-1}\left(J_{\mathcal{G}(i)}\otimes I_m\right)S^{-1}\ \mathsf{vec}\left[C^\top\check{Y}L^{-1}H_n\sqrt{G}\right] = \\
= \frac{1}{\nu_i}\mathsf{vec}\left[\left(Q_B^\top C^\top\check{Y}L^{-1}H_n\sqrt{G}Q_A\right)\odot M\right]^\top &\left(Q_A^\top J_{\mathcal{G}(i)}Q_A\otimes I_m\right)\ \mathsf{vec}\left[\left(Q_B^\top C^\top\check{Y}L^{-1}H_n\sqrt{G}Q_A\right)\odot M\right] = \\
= \frac{1}{\nu_i}\mathsf{vec}\left[\left(\left(Q_B^\top C^\top\check{Y}L^{-1}H_n\sqrt{G}Q_A\right)\odot M\right)Q_A^\top J_{\mathcal{G}(i)}\right]^\top &\ \mathsf{vec}\left[\left(\left(Q_B^\top C^\top\check{Y}L^{-1}H_n\sqrt{G}Q_A\right)\odot M\right)Q_A^\top J_{\mathcal{G}(i)}\right],
\end{aligned} \tag{75}$$

where $M$ is a matricization as in the previous section. For practical purposes, let's expand one of the matrix products: $Q_B^\top C^\top\check{Y}L^{-1}H_n\sqrt{G}Q_A = Q_B^\top\sqrt{\Sigma}B^\top Y H_n L^{-T}L^{-1}\sqrt{G}Q_A = Q_B^\top\sqrt{\Sigma}Z\sqrt{G}Q_A$, where $Z = B^\top Y H_n\left(H_n^\top D H_n\right)^{-1}H_n$. Then,

$$\begin{aligned}
-\frac{\partial \mathsf{vec}\left[\check{Y}\right]^\top \hat{S}^{-1}\ \mathsf{vec}\left[\check{Y}\right]}{\partial \nu_i} &= \frac{1}{\nu_i}\mathrm{tr}\left\{\underbrace{\left(\left(Q_B^\top\sqrt{\Sigma}Z\sqrt{G}Q_A\right)\odot M\right)Q_A^\top}_{\Lambda_\nu} J_{\mathcal{G}(i)}Q_A\left(\left(Q_B^\top\sqrt{\Sigma}Z\sqrt{G}Q_A\right)\odot M\right)^\top\right\} = \\
&= \frac{1}{\nu_i}\mathrm{tr}\left\{\Lambda_\nu^\top\Lambda_\nu J_{\mathcal{G}(i)}\right\},
\end{aligned} \tag{76}$$

or, in a vector form:

$$-\frac{\partial \mathsf{vec}\left[\check{Y}\right]^\top \hat{S}^{-1}\ \mathsf{vec}\left[\check{Y}\right]}{\partial \boldsymbol{\nu}} = -\frac{1}{\boldsymbol{\nu}} \odot \left(\mathcal{J}_{\mathcal{G}}\mathtt{diag}[\Lambda_\nu\Lambda_\nu^\top]\right). \tag{77}$$

##### 4.6.2.2.2. Derivative of the determinant term

$$\begin{aligned}
\frac{\partial \ln\det\hat{S}}{\partial \nu_i} = \frac{\partial \ln\det S}{\partial \nu_i} &= \frac{1}{\sqrt{\nu_i}}\mathrm{tr}\left\{S^{-1}\left(A^T L^{-1}H_n J_{\mathcal{G}(i)}\right)\otimes(C^\top C)\right\} = \\
&= \frac{1}{\sqrt{\nu_i}}\mathrm{tr}\left\{(D_A\otimes D_B + I_{nm})^{-1}\left(Q_A^\top A^T L^{-1}H_n J_{\mathcal{G}(i)}Q_A\right)\otimes D_B\right\}.
\end{aligned} \tag{78}$$

If we denote $\left(Q_A^\top\sqrt{G}H_n^\top\left(H_n D H_n^\top\right)^{-1}H_n\right)\odot Q_A^\top$ as $Z_\nu$, $\mathtt{diag}\left[(D_A\otimes D_B + I_{nm})^{-1}\right]$ as $\boldsymbol{s}$ and $\mathtt{diag}[D_B]$ as $\boldsymbol{d_B}$, then the derivative can be rewritten in the vector form:

$$\frac{\partial \ln\det\hat{S}}{\partial \boldsymbol{\nu}} = \frac{1}{\sqrt{\boldsymbol{\nu}}} \odot \left(\mathcal{J}_{\mathcal{G}}(Z_\nu\otimes\boldsymbol{d_B})\boldsymbol{s}\right). \tag{79}$$

### 4.6.3. A note on the identifiability of $\Sigma$ and $G$

In a general case, parametrized matrices $\Sigma$ and $G$ are not identifiable, see that $X \sim \mathcal{MN}(0, \Sigma, G)$ is the same random variable as $X \sim \mathcal{MN}(0, c\Sigma, \frac{1}{c}G)$, where $c$ is any non-zero constant. That means that unless we impose some constraints, the scale of estimates of $\Sigma$ and $G$ can be arbitrary. In practice, this causes no problem for an optimization method, however, if we seek to study asymptotic properties of parameter estimates, it becomes an issue due to the fact that the Fisher information matrix (FIM) is not positive-definite. The problem can be dealt with by setting $\tau_j$ to any positive constant value. In MARADONER, we set it to $\frac{\sigma_j}{4}$, where $j$ is index of a group with the smallest $\sigma_i$.

### 4.6.4. Fisher information matrix

Fisher information matrix (FIM) $\mathfrak{I}$ for a multivariate-normal r.v. $X \sim \mathcal{N}(0, S)$ has a known general form:

$$\mathfrak{I}_{i,j} = \frac{1}{2}\operatorname{tr}\left\{\hat{S}^{-1}\frac{\partial \hat{S}}{\partial \theta_i}\hat{S}^{-1}\frac{\partial \hat{S}}{\partial \theta_j}\right\}. \tag{80}$$

For Equation 55, $\hat{S} = (L^{-1}H_n G H_n^\top L^{-T}) \otimes (B\Sigma B^\top) + I_{(p-1)(n-1)} = (AA^\top) \otimes (CC^\top) + I_{(p-1)(n-1)}$.

$$\frac{\partial \hat{S}}{\partial \tau_i} = (AA^\top) \otimes (BJ_iB^\top) = \left[A \otimes \left(B\sqrt{\Sigma}\right)\right]\left[A^\top \otimes \left(\sqrt{\Sigma^{-1}}J_iB^\top\right)\right] = \frac{1}{\sqrt{\tau_i}}[A \otimes C][A^\top \otimes (J_iB^\top)], \tag{81}$$

$$\begin{aligned}\frac{\partial \hat{S}}{\partial \nu_i} &= (L^{-1}H_nJ_iH_n^\top L^{-T}) \otimes (CC^\top) = \frac{1}{\sqrt{\nu_i}}\left[\left(L^{-1}H_n\sqrt{G}\right) \otimes C\right]\left[(J_iH_n^\top L^{-T}) \otimes C^\top\right] = \\ &= \frac{1}{\sqrt{\nu_i}}[A \otimes C]\left[(J_iH_n^\top L^{-T}) \otimes C^\top\right].\end{aligned} \tag{82}$$

The same trick due to Equation 2 is applicable:

$$\begin{aligned}\hat{S}^{-1}\frac{\partial \hat{S}}{\partial \tau_i} &= \frac{1}{\sqrt{\tau_i}}[A \otimes C]S^{-1}[A^\top \otimes (J_iB^\top)] = \frac{1}{\sqrt{\tau_i}}[(AQ_A) \otimes (CQ_B)](D_A \otimes D_B + I_{nm})^{-1}\left[(Q_A^\top A^\top) \otimes \right. \\ &\left.\otimes\, (Q_B^\top J_iB^\top)\right] = \frac{1}{\sqrt{\tau_i}}\underbrace{[(AQ_A) \otimes (CQ_B)](D_A \otimes D_B + I_{nm})^{-1}}_{V}\left[(Q_A^\top A^\top) \otimes (Q_B^\top J_iB^\top)\right] = \\ &= \frac{1}{\sqrt{\tau_i}}V\left[(Q_A^\top A^\top) \otimes (Q_B^\top J_iB^\top)\right],\end{aligned} \tag{83}$$

$$\hat{S}^{-1}\frac{\partial \hat{S}}{\partial \nu_i} = \frac{1}{\sqrt{\nu_i}}V\left(Q_A^\top J_{\mathcal{G}(i)}H_n^\top L^{-T}\right) \otimes (Q_B^\top C^\top). \tag{84}$$

Let's consider 3 cases:
1. Both parameters are in $\boldsymbol{\tau}$ – $\mathfrak{I}_{\boldsymbol{\tau}}$;
2. Both parameters are in $\boldsymbol{\nu}$ – $\mathfrak{I}_{\boldsymbol{\nu}}$;
3. Mixed case: one parameter in $\boldsymbol{\tau}$ and the other is in $\boldsymbol{\nu}$ – $\mathfrak{I}_{\boldsymbol{\tau\nu}}$.

$$\mathfrak{I}(\boldsymbol{\tau}, \boldsymbol{\nu}) =$$

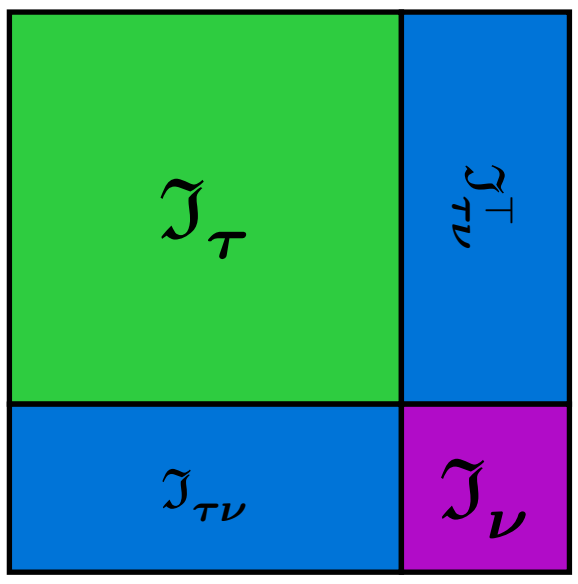

. (85)

#### 4.6.4.1. Both parameters are in $\boldsymbol{\tau}$

$$\operatorname{tr}\left\{\hat{S}^{-1}\frac{\partial \hat{S}}{\partial \tau_i}\hat{S}^{-1}\frac{\partial \hat{S}}{\partial \tau_j}\right\} = \frac{1}{\sqrt{\tau_i\tau_j}}\operatorname{tr}\left\{V\left[(Q_A^\top A^\top) \otimes (Q_B^\top J_iB^\top)\right]V\left[(Q_A^\top A^\top) \otimes (Q_B^\top J_jB^\top)\right]\right\}, \tag{86}$$

where $\left[(Q_A^\top A^\top) \otimes (Q_B^\top J_iB^\top)\right]V = [D_A \otimes (Q_B^\top J_iB^\top CQ_B)](D_A \otimes D_B + I_{nm})^{-1} = \mathrm{T}_i$. Then:

$$\operatorname{tr}\left\{\hat{S}^{-1}\frac{\partial \hat{S}}{\partial \tau_i}\hat{S}^{-1}\frac{\partial \hat{S}}{\partial \tau_j}\right\} = \frac{1}{\sqrt{\tau_i\tau_j}}\operatorname{tr}\{\mathrm{T}_i\mathrm{T}_j\}. \tag{87}$$

Computing Kronecker products naively is incompatible with most modern PC's RAM capacities. However, for $\hat{S}^{-1}\frac{\partial \hat{S}}{\partial \tau_i}$, Kronecker product results in a block-diagonal matrix, making $\operatorname{tr}\left\{\hat{S}^{-1}\frac{\partial \hat{S}}{\partial \tau_i}\hat{S}^{-1}\frac{\partial \hat{S}}{\partial \tau_j}\right\}$ computable in a much more memory-efficient manner. Let's denote $k$-th block of $T_i$ as $T_i^k$, of $(D_A \otimes D_B + I_{nm})^{-1}$ as $S^k$, $\alpha_k$ as $k$-th diagonal element of $D_A$ and $(Q_B^\top J_iB^\top CQ_B)$ as $F_i$ for brevity:

$$(D_A \otimes D_B + I_{nm})^{-1} = S = \left[\begin{array}{c|c|c} S^1 & \dots & 0 \\ \hline \vdots & \ddots & \vdots \\ \hline 0 & \dots & S^n \end{array}\right], S^i = \begin{pmatrix} s_{(i-1)m} & 0 & \dots & 0 \\ 0 & s_{(i-1)m+1} & \dots & 0 \\ \vdots & \vdots & \ddots & \vdots \\ 0 & 0 & \dots & s_{im-1} \end{pmatrix}, D_A \otimes F_i = \left[\begin{array}{c|c|c} \alpha_1F_i & \dots & 0 \\ \hline \vdots & \ddots & \vdots \\ \hline 0 & \dots & \alpha_nF_i \end{array}\right], \tag{88}$$

so

$$T_i = \begin{bmatrix} \alpha_1 F_i & \dots & 0 \\ \vdots & \ddots & \vdots \\ 0 & \dots & \alpha_n F_i \end{bmatrix} \begin{bmatrix} S^1 & \dots & 0 \\ \vdots & \ddots & \vdots \\ 0 & \dots & S^n \end{bmatrix} = \begin{bmatrix} \alpha_1 F_i S^1 & \dots & 0 \\ \vdots & \ddots & \vdots \\ 0 & \dots & \alpha_n F_i S^n \end{bmatrix} = \begin{bmatrix} T_i^1 & \dots & 0 \\ \vdots & \ddots & \vdots \\ 0 & \dots & T_i^n \end{bmatrix}. \tag{89}$$

Then:

$$\begin{aligned} \operatorname{tr}\{T_i T_j\} &= \operatorname{tr}\left\{ \begin{bmatrix} T_i^1 & \dots & 0 \\ \vdots & \ddots & \vdots \\ 0 & \dots & T_i^n \end{bmatrix} \begin{bmatrix} T_j^1 & \dots & 0 \\ \vdots & \ddots & \vdots \\ 0 & \dots & T_j^n \end{bmatrix} \right\} = \sum_{k=1}^{n} \operatorname{tr}\{T_i^k T_j^k\} = \sum_{k=1}^{n} \alpha_k^2 \operatorname{tr}\{F_i S_k F_j F_k\} = \\ &= \sum_{k=1}^{n} \alpha_k^2 \operatorname{tr}\{(Q_B^\top J_i B^\top C Q_B) S_k (Q_B^\top J_j B^\top C Q_B) S_k\} = \sum_{k=1}^{n} \alpha_k^2 \operatorname{tr}\left\{ Q_B^\top J_i \underbrace{B^\top C Q_B S_k Q_B^\top}_{\Lambda_k} J_j B^\top C Q_B S_k \right\} = \\ &= \sum_{k=1}^{n} \alpha_k^2 \operatorname{tr}\{\Lambda_k J_i \Lambda_k J_j\} = \sum_{k=1}^{n} \alpha_k^2 \left[\Lambda_k \odot \Lambda_k^\top\right]_{[ij]}. \end{aligned} \tag{90}$$

Hence,

$$\mathfrak{I}_{\boldsymbol{\tau}} = \frac{1}{2} \left( \sum_{k=1}^{n} \alpha_k^2 \Lambda_k \odot \Lambda_k^\top \right) \odot \left( \sqrt{\boldsymbol{\tau}^{-1}} \sqrt{\boldsymbol{\tau}^{-1}}^\top \right). \tag{91}$$

#### 4.6.4.2. Both parameters are in $\nu$

$$\operatorname{tr}\left\{ \hat{S}^{-1} \frac{\partial \hat{S}}{\partial \nu_i} \hat{S}^{-1} \frac{\partial \hat{S}}{\partial \nu_j} \right\} = \frac{1}{\sqrt{\nu_i \nu_j}} \operatorname{tr}\{\mathrm{N}_i \mathrm{N}_j\}, \tag{92}$$

where $\mathrm{N}_i = \left[\left(Q_A^\top J_{\mathcal{G}(i)} H_n^\top L^{-\top} A Q_A\right) \otimes D_B\right] (D_A \otimes D_B + I_{nm})^{-1}$. Unfortunately, $\mathrm{N}_i$ is not a block-diagonal matrix like $\mathrm{T}_i$ due to the diagonal matrix $D_B$ being at the latter position in the first Kronecker product. Kronecker products are also non-commutative, so we can't switch $D_B$ with $Q_A^\top J_{\mathcal{G}(i)} H_n^\top L^{-\top} A Q_A$ and call it a day. However, for any Kronecker product $A \otimes B$, where $A$ and $B$ are square matrices o sizes $n$ and $m$ respectively, there always exist a special type of permutation matrix called "commutation matrix" $\mathcal{C}_{(n,m)}$ such that $A \otimes B = \mathcal{C}_{(n,m)}^\top (B \otimes A) \mathcal{C}_{(n,m)}$. Then:

$$\begin{aligned} \operatorname{tr}\{N_i N_j\} &= \operatorname{tr}\Big\{ \mathcal{C}_{(n,m)}^\top \left[D_B \otimes \left(Q_A^\top J_{\mathcal{G}(i)} H_n^\top L^{-\top} A Q_A\right)\right] \mathcal{C}_{(n,m)} (D_A \otimes D_B + I_{nm})^{-1} \times \\ &\times \mathcal{C}_{(n,m)}^\top \left[D_B \otimes \left(Q_A^\top J_{\mathcal{G}(j)} H_n^\top L^{-\top} A Q_A\right)\right] \mathcal{C}_{(n,m)} (D_A \otimes D_B + I_{nm})^{-1} \Big\} = \\ &= \operatorname{tr}\Big\{ \left[D_B \otimes \left(Q_A^\top J_{\mathcal{G}(i)} H_n^\top L^{-\top} A Q_A\right)\right] \mathcal{C}_{(n,m)} (D_A \otimes D_B + I_{nm})^{-1} \mathcal{C}_{(n,m)}^\top \times \\ &\times \left[D_B \otimes \left(Q_A^\top J_{\mathcal{G}(j)} H_n^\top L^{-\top} A Q_A\right)\right] \mathcal{C}_{(n,m)} (D_A \otimes D_B + I_{nm})^{-1} \mathcal{C}_{(n,m)}^\top \Big\}. \end{aligned} \tag{93}$$

Let's inspect matrix $\mathcal{C}_{(n,m)} (D_A \otimes D_B + I_{nm})^{-1} \mathcal{C}_{(n,m)}^\top$.

**Property 4.6.4.2.1** (On the permutation rule): Let $S$ be a diagonal matrix $S = \texttt{diag}\{s_i\}_{i=1}^{n*m}$ and $\mathcal{C}_{(n,m)}$ be a commutation matrix. Then, matrix $\hat{S}$

$$\hat{S} = \mathcal{C}_{(n,m)} S \mathcal{C}_{(n,m)}^\top, \tag{94}$$

1. Is a diagonal matrix too.
2. Its diagonal elements are extracted from $\{s_i\}_{\{i=1\}}^{\{nm\}}$ with the permutation rule

$$\pi(i) = n(i \% m) + \left\lfloor \frac{i}{m} \right\rfloor. \tag{95}$$

*Proof*: Let $\boldsymbol{e_k}$ be $k$-th standard basis vector. Then, if we denote the permutation rule given by $\mathcal{C}_{(n,m)}$ as $\sigma(i)$ and, consequently, $\sigma^{-1}(i)$ for $\mathcal{C}_{(n,m)}^\top$, then $\mathcal{C}_{(n,m)}^\top \boldsymbol{e_k} = \boldsymbol{e_{\sigma^{-1}(k)}}$, $S\boldsymbol{e_{\sigma^{-1}(k)}} = s_{\sigma^{-1}(k)} \boldsymbol{e_{\sigma^{-1}(k)}}$ and finally $s_{\sigma^{-1}(k)} \mathcal{C}_{(n,m)} \boldsymbol{e_{\sigma^{-1}(k)}} = s_{\sigma^{-1}(k)} \boldsymbol{e_k}$. So the $\hat{S}$ is diagonal indeed and its diagonal elements are given by the $\sigma^{-1}$ permutation of diagonal elements of $S$.

What's the $\sigma$? By the definition of the commutation matrix, for any matrix $A$ of shape $n \times m$ $\mathcal{C}_{(n,m)}$ $\texttt{vec}[A] = \texttt{vec}[A^\top]$. The position of element $a_{i,j}$ in $\texttt{vec}[A]$ is $i + nj$. The position of the same element in $\texttt{vec}[A^\top]$ is $j + mi$. So $\sigma(k) = j + mi = \lfloor \frac{k}{n} \rfloor + m(k\%n)$. Therefore, $\pi(i) = \sigma^{-1}(i) = n(i\%m) + \lfloor \frac{i}{m} \rfloor$.

■

Hence, if $\hat{\mathrm{N}}_i = \left[D_B \otimes \left(Q_A^\top J_{\mathcal{G}(j)} H_n^\top L^{-\top} A Q_A\right)\right] \mathcal{C}_{(n,m)} (D_A \otimes D_B + I_{nm})^{-1} \mathcal{C}_{(n,m)}^\top$, then

$$\mathrm{tr}\left\{\hat{S}^{-1}\frac{\partial \hat{S}}{\partial \nu_i}\hat{S}^{-1}\frac{\partial \hat{S}}{\partial \nu_j}\right\} = \frac{1}{\sqrt{\nu_i \nu_j}}\mathrm{tr}\left\{\mathrm{N}_i \mathrm{N}_j\right\} = \frac{1}{\sqrt{\nu_i \nu_j}}\mathrm{tr}\left\{\hat{\mathrm{N}}_i \hat{\mathrm{N}}_j\right\}, \tag{96}$$

and, like in the previous case, if we denote $\beta_k$ as $k$-th element of the diagonal matrix $D_B$, and $\hat{S}_k$ now the $k$-th block of matrix $\mathcal{C}_{(n,m)}(D_A \otimes D_B + I_{nm})^{-1}\mathcal{C}_{(n,m)}^\top$ of size $n \times n$, then

$$\begin{aligned} \mathrm{tr}\left\{\hat{\mathrm{N}}_i \hat{\mathrm{N}}_j\right\} &= \sum_{k=1}^m \beta_k \,\mathrm{tr}\left\{ Q_A^\top J_{\mathcal{G}(j)} \underbrace{H_n^\top L^{-\top} A Q_A \hat{S}_k Q_A^\top}_{\Gamma_k} J_{\mathcal{G}(i)} H_n^\top L^{-\top} L^{-T} A Q_A \hat{S}_k \right\} = \\ &= \sum_{k=1}^m \beta_k \,\mathrm{tr}\left\{\Gamma_k J_{\mathcal{G}(i)} \Gamma_k J_{\mathcal{G}(j)}\right\} = \sum_{k=1}^m \beta_k \sum_{v \in \mathcal{G}(i)} \sum_{p \in \mathcal{G}(j)} \left[\Gamma_k \odot \Gamma_k^\top\right]_{v,p}, \end{aligned} \tag{97}$$

and, finally,

$$\mathfrak{I}_{\boldsymbol{\nu}} = \frac{1}{2}\mathcal{G}\left(\sum_{k=1}^m \beta_k^2 \Gamma_k \odot \Gamma_k^\top\right)\mathcal{G}^\top \odot \left(\sqrt{\boldsymbol{\nu}^{-1}}\sqrt{\boldsymbol{\nu}^{-1}}^\top\right), \tag{98}$$

where $\mathcal{G}$ is a $g \times n$ matrix of loadings of samples onto groups.

#### 4.6.4.3. Mixed case

In this section are abusing notation: we shall use zero-based vector numberings, but $\boldsymbol{e_k}$ is $k-1$-th basis vector. That's done to avoid writing in extra $\pm 1$ in formulae to come.

$$\mathrm{tr}\left\{\hat{S}^{-1}\frac{\partial \hat{S}}{\partial \tau_i}\hat{S}^{-1}\frac{\partial \hat{S}}{\partial \nu_j}\right\} = \frac{1}{\sqrt{\tau_i \nu_j}}\mathrm{tr}\left\{\mathrm{T}_i \mathrm{N}_j\right\} \tag{99}$$

This is the toughest case as no block-diagonal representation is available for the $T_i N_j$ product. There is not much we can do about this trace, but at the same time we can't compute it straightforwardly due to tremendous memory requirements of the naive Kronecker product. Instead, see that for any square $n \times n$ matrix $X$

$$\mathrm{tr}(X) = \sum_{k=1}^n \boldsymbol{e_k}^\top A \boldsymbol{e_k}, \tag{100}$$

i.e. it can be represented as a sum of quadratic forms. This can be used to obtain memory-efficient algorithm to compute $\mathrm{tr}\{\mathrm{T}_i \mathrm{N}_j\}$:

$$\mathrm{tr}\left\{\mathrm{T}_i \mathrm{N}_j\right\} = \sum_{k=1}^{nm} \boldsymbol{e_k}^\top \mathrm{T}_i \mathrm{N}_j \boldsymbol{e_k} = \sum_{k=1}^{nm} (\mathrm{T}_i \boldsymbol{e_k})^\top \mathrm{N}_j \boldsymbol{e_k}, \tag{101}$$

where

$$\mathrm{T}_i \boldsymbol{e_k} = \left[D_A \otimes \left(Q_B^\top J_i \underbrace{B^\top C Q_B}_{V}\right)\right] \underbrace{(D_A \otimes D_B + I_{nm})^{-1} \boldsymbol{e_k}}_{\boldsymbol{s_k}} = \texttt{vec}[Q_B^\top J_i V \ \texttt{mat}[\boldsymbol{s_k}] D_A], \tag{102}$$

$$\mathrm{N}_j \boldsymbol{e_k} = \left[\left(Q_A^\top J_{\mathcal{G}(j)} \underbrace{H_n^\top L^{-\top} A Q_A}_{K}\right) \otimes D_B\right] \boldsymbol{s_k} = \texttt{vec}\left[D_B \ \texttt{mat}[\boldsymbol{s_k}] K^\top J_{\mathcal{G}(j)} Q_A\right]. \tag{103}$$

Here, matricization of $\boldsymbol{s_k}$ is just a zero matrix with $\left(k\%m, \lfloor \frac{k}{m} \rfloor\right)$ set to $k$-th element of $(D_A \otimes D_B + I_{nm})^{-1}$, denoted as $s_k$. Or, in a matrix form, $\texttt{mat}[s_k] = s_k \boldsymbol{e}^m_{\boldsymbol{k\%m}} \boldsymbol{e}^n_{\lfloor \frac{k}{m} \rfloor}{}^\top$, where superscript denotes the length of the basis vector explicitly. Note that

$$\texttt{mat}[\boldsymbol{s_k}]D_A = \alpha_{\lfloor \frac{k}{m} \rfloor} s_k \boldsymbol{e^m_{k\%m}} \boldsymbol{e^n_{\lfloor \frac{k}{m} \rfloor}}^\top \tag{104}$$

and

$$D_B\ \texttt{mat}[\boldsymbol{s_k}] = \beta_{k\%m} s_k \boldsymbol{e^m_{k\%m}} \boldsymbol{e^n_{\lfloor \frac{k}{m} \rfloor}}^\top \tag{105}$$

.

Let's now inspect terms $K^\top J_{\mathcal{G}(i)} Q_A$ and $Q_B^\top J_i V$. As $J_i$ and $J_{\mathcal{G}(i)}$ idepmotent matrices, $K^\top J_{\mathcal{G}(i)} Q_A = K^\top J_{\mathcal{G}(i)} J_{\mathcal{G}(i)} Q_A$ and $Q_B^\top J_i V = Q_B^\top J_i J_i V$ and, if we denote $i$-th of $K$ row as $\boldsymbol{k_i}$, $i$-th row of $Q_A$ as $\boldsymbol{a_i}$, $i$-th row of $Q_B$ as $\boldsymbol{b_i}$ and $i$-th row of $V$ as $\boldsymbol{v_i}$, then

$$Q_B^\top J_i V = \boldsymbol{b}_i \boldsymbol{v}_i^\top,\ K^T J_{\mathcal{G}(j)} Q_A = \sum_{l \in \mathcal{G}(j)} \boldsymbol{k_l} \boldsymbol{a_l}^\top. \tag{106}$$

Next, let's combine those terms together:

$$\texttt{mat}[\mathrm{T}_i \boldsymbol{e_k}] = Q_B^\top J_i V\ \texttt{mat}[\boldsymbol{s_k}] D_A = a_{\lfloor \frac{k}{m} \rfloor} s_k \boldsymbol{b_i} \boldsymbol{v}_i^\top \boldsymbol{e^m_{k\%m}} \boldsymbol{e^n_{\lfloor \frac{k}{m} \rfloor}}^\top = v^i_{k\%m} \alpha_{\lfloor \frac{k}{m} \rfloor} s_k \boldsymbol{b_i} \boldsymbol{e^n_{\lfloor \frac{k}{m} \rfloor}}^\top, \tag{107}$$

$$\begin{aligned} \texttt{mat}\left[\mathrm{N}_j \boldsymbol{e_k}\right] = D_B\ \texttt{mat}[\boldsymbol{s_k}] K^T J_{\mathcal{G}(j)} Q_A = D_B\ \texttt{mat}[\boldsymbol{s_k}] = \beta_{k\%m} s_k \boldsymbol{e^m_{k\%m}} \boldsymbol{e^n_{\lfloor \frac{k}{m} \rfloor}}^\top \sum_{i \in \mathcal{G}(j)} \boldsymbol{k_i} \boldsymbol{a}_i^\top = \\ = \beta_{k\%m} s_k \boldsymbol{e^m_{k\%m}} \sum_{i \in \mathcal{G}(j)} k^i_{\lfloor \frac{k}{m} \rfloor} \boldsymbol{a_i}^\top. \end{aligned} \tag{108}$$

First, note that in $\texttt{mat}[\mathrm{T}_i \boldsymbol{e_k}]$ term $\boldsymbol{b_i} \boldsymbol{e^n_{\lfloor \frac{k}{m} \rfloor}}^\top$ results in a zero matrix with $\lfloor \frac{k}{m} \rfloor$-th column set to $\boldsymbol{b_i}$, and in $\texttt{mat}\left[\mathrm{N}_j \boldsymbol{e_k}\right]$ term $\boldsymbol{e^m_{k\%m}} \boldsymbol{a_i}^\top$ result in a zero matrix with $k\%m$-th row set to $\boldsymbol{a_i}$. Then, $\mathrm{T}_i \boldsymbol{e_k}$ is zero everywhere except for indices at a range $\left[m\lfloor \frac{k}{m} \rfloor, m\lfloor \frac{k}{m} \rfloor + m\right)$. The other vector, $\mathrm{N}_j \boldsymbol{e_k}$, however, attains a different sparse structure: its non-zero elements are located at indices $\{k\%m + cm\}_{c=0}^{n-1}$. That means that in dot product $(\mathrm{T}_i \boldsymbol{e_k})^\top (\mathrm{N}_j \boldsymbol{e_k})$ only the terms that lie in the intersection of the interval and the set do contribute:

$$m\left\lfloor \frac{k}{m} \right\rfloor \le k\%m + cm < m\left(\left\lfloor \frac{k}{m} \right\rfloor + 1\right), \tag{109}$$

which is possible only for a single value of $c$ for a particular $k$.

$$\begin{aligned} m\left\lfloor \frac{k}{m} \right\rfloor \le \overbrace{k - m\left\lfloor \frac{k}{m} \right\rfloor}^{k\%m} + cm < m\left(\left\lfloor \frac{k}{m} \right\rfloor + 1\right) \leftrightarrow 2\left\lfloor \frac{k}{m} \right\rfloor - \frac{k}{m} \le c < 2\left\lfloor \frac{k}{m} \right\rfloor - \frac{k}{m} + 1 \leftrightarrow \\ \left\lfloor \frac{k}{m} \right\rfloor - \underbrace{\left(\frac{k}{m} - \left\lfloor \frac{k}{m} \right\rfloor\right)}_{\{\frac{k}{m}\} \le 1} \le c < \left\lfloor \frac{k}{m} \right\rfloor + \underbrace{1 - \left(\frac{k}{m} - \left\lfloor \frac{k}{m} \right\rfloor\right)}_{1 - \{\frac{k}{m}\} \le 1}, \end{aligned} \tag{110}$$

or, graphically,

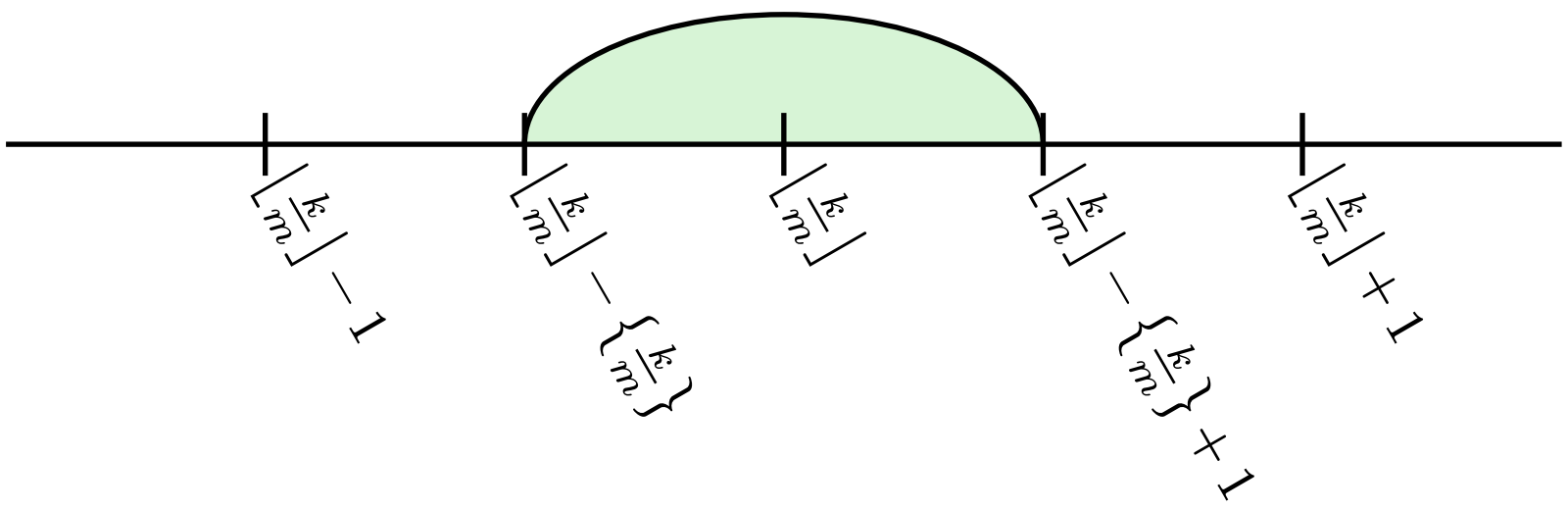

where $\lfloor \frac{k}{m} \rfloor$ is the only integer in the interval $\to c = \lfloor \frac{k}{m} \rfloor$. Hence, we need only the $c = \lfloor \frac{k}{m} \rfloor$-th element of the $\boldsymbol{a_i}$ and $k\%m + cm - m\lfloor \frac{k}{m} \rfloor = k\%m$-th element of the $\boldsymbol{b_i}$. Finally,

$$(\mathrm{T}_i \boldsymbol{e_k})^\top (\mathrm{N}_j \boldsymbol{e_k}) = s_k^2 v^i_{k\%m} \alpha_{\lfloor \frac{k}{m} \rfloor} \beta_{k\%m} b^i_{k\%m} \sum_{l \in \mathcal{G}(j)} k^l_{\lfloor \frac{k}{m} \rfloor} a^l_{\lfloor \frac{k}{m} \rfloor}. \tag{111}$$

Therefore, to compute $\mathtt{tr}\{\mathrm{T}_i \mathrm{N}_j\}$ we need only $O(mn)$ operations:

$$[\mathfrak{I}_{\boldsymbol{\tau\nu}}]_{i,j} = \frac{1}{2}\frac{1}{\sqrt{\tau_i\nu_j}}\operatorname{tr}\{\mathrm{T}_i\mathrm{N}_j\} = \frac{1}{2}\frac{1}{\sqrt{\tau_i\nu_j}}\sum_{k=1}^{nm}\zeta_k\psi_k^i\sum_{l\in\mathcal{G}(j)}\hat{\theta}_k^l = \frac{1}{2}\frac{1}{\sqrt{\tau_i\nu_j}}\sum_{k=1}^{nm}\zeta_k\psi_k^i\theta_k^j \tag{112}$$

where $\zeta_k = s_k^2\alpha_{\lfloor\frac{k}{m}\rfloor}\beta_{k\%m}$, $\psi_k^i = v_{k\%m}^i b_{k\%m}^i = [V\odot Q_B]_{[i,k\%m]}$, $\hat{\theta}_k^l = k_{\lfloor\frac{k}{m}\rfloor}^l a_{\lfloor\frac{k}{m}\rfloor}^l = [K\odot Q_A]_{[l,\lfloor\frac{k}{m}\rfloor]}$, $\theta_k^j = [(K\odot Q_A)\mathcal{G}]_{[j,\lfloor\frac{k}{m}\rfloor]}$. In the principle, we can construct matrices $\Psi$, $\mathrm{Z}$, $\Theta$ by appropriately repeating $\theta$, $\zeta$ and $\psi$ terms Equation 112:

$$\mathfrak{I}_{\tau\nu} = \frac{1}{2}(\Psi\mathrm{Z}\Theta)\odot\frac{1}{\sqrt{\boldsymbol{\tau\nu}^\top}} \tag{113}$$

However, doing so will require storing $\Psi$ and $\Theta$ ($\mathrm{Z}$ being diagonal requires just $nm$ storage), each having $n^2m$ and $nm^2$ entries, which might be infeasible for some PCs and datasets. Therefore we are sticking to the element-wise representation from Equation 112 for computing $\mathfrak{I}_{\boldsymbol{\tau\nu}}$.

## 4.7. Estimating $\mu_m$

We shall proceed similarly to Section 4.5, but this time for a $\tilde{Y} = (Y - H_p\boldsymbol{\mu_p}\mathbf{1_n}^\top)\sqrt{D}^{-1}$:

$$\tilde{Y} = BU + \mathrm{E},\ U\sim\mathcal{MN}\left(\boldsymbol{\mu_m}\Bigg(\underbrace{\sqrt{D}^{-1}\mathbf{1_n}}_{\sqrt{\boldsymbol{d}^{-1}}}\Bigg)^\top, \Sigma, \underbrace{\sqrt{D}^{-1}G\sqrt{D}^{-1}}_{\hat{G}}\right), \mathrm{E}\sim\mathcal{MN}(0, I_{p-1}, I_n), \tag{114}$$

where $\hat{G}$ is a diagonal matrix with non-zero entries $\hat{g}_i = \frac{\nu_i}{\sigma_i}$.

### 4.7.1. Calculating log-likelihood

Then, the log-likelihood is:

$$\begin{gathered}
l(\tilde{Y}|\boldsymbol{\mu_m}) = \operatorname{vec}\left[\tilde{Y} - B\boldsymbol{\mu_m}\sqrt{\boldsymbol{d}^{-1}}^\top\right]^\top\left(\hat{G}\otimes(B\Sigma B^\top) + I_{n(p-1)}\right)^{-1}\operatorname{vec}\left[\tilde{Y} - B\boldsymbol{\mu_m}\sqrt{\boldsymbol{d}^{-1}}^\top\right] = \\
= \operatorname{tr}\left\{\left(\tilde{Y} - B\boldsymbol{\mu_m}\sqrt{\boldsymbol{d}^{-1}}^\top\right)\left(\tilde{Y} - B\boldsymbol{\mu_m}\sqrt{\boldsymbol{d}^{-1}}^\top\right)^\top\right\} - \operatorname{vec}\left[\tilde{Y} - B\boldsymbol{\mu_m}\sqrt{\boldsymbol{d}^{-1}}^\top\right]^\top\left(\sqrt{\hat{G}}\otimes\left(B\sqrt{\Sigma}\right)\right)\times \\
\times\Bigg(\underbrace{\hat{G}\otimes\left(\sqrt{\Sigma}B^\top B\sqrt{\Sigma}\right) + I_{nm}}_{S}\Bigg)^{-1}\left(\sqrt{\hat{G}}\otimes\left(\sqrt{\Sigma}B^\top\right)\right)\operatorname{vec}\left[\tilde{Y} - B\boldsymbol{\mu_m}\sqrt{\boldsymbol{d}^{-1}}^\top\right] = \operatorname{tr}\left\{\left(\tilde{Y} - B\boldsymbol{\mu_m}\sqrt{\boldsymbol{d}^{-1}}^\top\right)\times\right. \\
\left.\times\left(\tilde{Y} - B\boldsymbol{\mu_m}\sqrt{\boldsymbol{d}^{-1}}^\top\right)^\top\right\} - \operatorname{vec}\left[\sqrt{\Sigma}B^\top\left(\tilde{Y} - B\boldsymbol{\mu_m}\sqrt{\boldsymbol{d}^{-1}}^\top\right)\sqrt{\hat{G}}\right]^\top S^{-1}\operatorname{vec}\left[\sqrt{\Sigma}B^\top\left(\tilde{Y} - B\boldsymbol{\mu_m}\sqrt{\boldsymbol{d}^{-1}}^\top\right)\sqrt{\hat{G}}\right].
\end{gathered} \tag{115}$$

Matrix $S$ has a block-diagonal structure ($n$ blocks of which $g$ blocks are unique), where each block looks like $\hat{g}_i\sqrt{\Sigma}B^\top B\sqrt{\Sigma} + I_p$.

If $Q_BD_BQ_B^\top$ is an eigendecomposition of $\sqrt{\Sigma}B^\top B\sqrt{\Sigma}$, then each diagonal block can be represented as $Q_B(\hat{g}_iD_B + I_m)Q_B^\top$. Also, we rename some products for brevity: $\sqrt{\boldsymbol{d}^{-1}}^\top\sqrt{\hat{G}} = \boldsymbol{r}^\top$, $\sqrt{\Sigma}B^\top\tilde{Y}\sqrt{\hat{G}} = \sqrt{\Sigma}B^\top(Y - H_p\boldsymbol{\mu_p}\mathbf{1_n}^\top)\sqrt{D}^{-1}\sqrt{\hat{G}} = \sqrt{\Sigma}B^\top(Y - H_p\boldsymbol{\mu_p}\mathbf{1_n}^\top)\sqrt{R} = \hat{Y}$ and $B^\top B = C$ Then, the latter term in $l(\tilde{Y}|\boldsymbol{\mu_m})$ can be represented as:

$$\begin{gathered}
\psi(\boldsymbol{\mu_m}) = \sum_{i=1}^{n}\operatorname{vec}\left[\left(\hat{Y} - \sqrt{\Sigma}C\boldsymbol{\mu_m}\boldsymbol{r}^\top\right)\boldsymbol{e_i}\right]^\top Q_B(\hat{g}_iD_B + I_m)^{-1}Q_B^\top\operatorname{vec}\left[\left(\hat{Y} - \sqrt{\Sigma}C\boldsymbol{\mu_m}\boldsymbol{r}^\top\right)\boldsymbol{e_i}\right] = \\
= \sum_{i=1}^{n}\left(\Bigg(\hat{Y} - \underbrace{\sqrt{\Sigma}C}_{A}\boldsymbol{\mu_m}\boldsymbol{r}^\top\Bigg)\boldsymbol{e_i}\right)^\top\underbrace{Q_B(\hat{g}_iD_B + I_m)^{-1}Q_B^\top}_{\hat{S}_i}\left(\hat{Y} - \sqrt{\Sigma}C\boldsymbol{\mu_m}\boldsymbol{r}^\top\right)\boldsymbol{e_i}. = \sum_{i=1}^{n}\left(\left(\hat{Y} - A\boldsymbol{\mu_m}\boldsymbol{r}^\top\right)\boldsymbol{e_i}\right)^\top\hat{S}^{-1}\times \\
\times\left(\left(\hat{Y} - A\boldsymbol{\mu_m}\boldsymbol{r}^\top\right)\boldsymbol{e_i}\right) = \sum_{i}^{n}\left(\hat{Y}\boldsymbol{e_i}\right)^\top\hat{S}_i^{-1}\hat{Y}\boldsymbol{e_i} + (A\boldsymbol{\mu_m}\boldsymbol{r}^\top\boldsymbol{e_i})^\top\hat{S}_i\left(A\boldsymbol{\mu_m}\sqrt{\boldsymbol{\nu}}^\top\boldsymbol{e_i}\right) - 2\left(\hat{Y}\boldsymbol{e_i}\right)^\top\hat{S}_i(A\boldsymbol{\mu_m}\boldsymbol{r}^\top\boldsymbol{e_i}) \approx \\
\approx \text{(dropping terms that do not depend on } \boldsymbol{\mu_m}) \approx \sum_{i}^{n}\underbrace{\boldsymbol{e_i}^\top\boldsymbol{r}}_{r_i}\boldsymbol{\mu_m}^\top A^\top\hat{S}_iA\boldsymbol{\mu_m}\boldsymbol{r}^\top\boldsymbol{e_i} - 2\boldsymbol{e_i}^\top\tilde{Y}^\top\hat{S}_iA\boldsymbol{\mu_m}\boldsymbol{r}^\top\boldsymbol{e_i} = \\
= \sum_{i}^{n}r_i^2\boldsymbol{\mu_m}^\top A^\top\hat{S}_iA\boldsymbol{\mu_m} - 2r_i\boldsymbol{e_i}^\top\hat{Y}^\top\hat{S}_iA\boldsymbol{\mu_m}.
\end{gathered} \tag{116}$$

Let's take derivative w.r.t to $\boldsymbol{\mu_m}$:

$$\begin{aligned}\frac{1}{2}\frac{\partial \psi\left(\hat{Y}|\boldsymbol{\mu_m}\right)}{\partial \boldsymbol{\mu_m}} &= \sum_i^n r_i^2 A^\top \hat{S}_i A\boldsymbol{\mu_m} - r_i A^\top \hat{S}_i \hat{Y}\boldsymbol{e_i} = A^\top\left(\sum_{i=1}^n r_i^2\hat{S}_i\right)A\boldsymbol{\mu_m} - A^\top\left(\sum_{i=1}^n r_i\hat{S}_i\hat{Y}\boldsymbol{e_i}\right) = \\ &= A^\top Q_B \left(\sum_{i=1}^n \frac{1}{\sigma_i}\left(D_B+\frac{\sigma_i}{\nu_i}I_m\right)^{-1}\right)Q_B^\top A\boldsymbol{\mu_m} - A^\top Q_B\left(\sum_{i=1}^n \frac{1}{\sqrt{\nu_i}}\left(D_B+\frac{\sigma_i}{\nu_i}I_m\right)^{-1}Q_B^\top\hat{Y}\boldsymbol{e_i}\right).\end{aligned} \tag{117}$$

Let's for brevity and ease of pratical implementation of forumalae in terms of vectorized operations, get rid of the summations by representing them in terms of various matrix products. Let's construct a matrix $\mathcal{S}$ of size $m \times n$ with $i$-th column equal to $\frac{1}{\sigma_i}$ vec$\left[\left(D_B+\frac{\sigma_i}{\nu_i}I_m\right)^{-1}\right]$. Then,

$$\frac{1}{2}\frac{\partial \psi(\boldsymbol{\mu_m})}{\partial \boldsymbol{\mu_m}} = \left(A^\top Q_B\right)\odot\left(\mathcal{S}\boldsymbol{1_n}\boldsymbol{1_m}^\top\right)^\top Q_B^\top A\boldsymbol{\mu_m} - A^\top Q_B\left(\mathcal{S}\odot\left(\boldsymbol{r^{-T}}\boldsymbol{1_n}\right)\right)\odot\left(Q_B^\top\hat{Y}\right)\boldsymbol{1_n}. \tag{118}$$

Next we set the derivative of $l$ to zero to solve for $\boldsymbol{\mu_m}$:

$$\begin{aligned}&\frac{\partial l\left(\hat{Y}|\boldsymbol{\mu_m}\right)}{\partial \boldsymbol{\mu_m}} = \frac{\partial}{\partial \boldsymbol{\mu_m}}\operatorname{tr}\left\{\left(\tilde{Y}-B\boldsymbol{\mu_m}\sqrt{\boldsymbol{d^{-1}}}^\top\right)\left(\tilde{Y}-B\boldsymbol{\mu_m}\sqrt{\boldsymbol{d^{-1}}}^\top\right)^\top\right\} - \frac{\partial \psi(\boldsymbol{\mu_m})}{\partial \boldsymbol{\mu_m}} = -\frac{\partial \psi(\boldsymbol{\mu_m})}{\partial \boldsymbol{\mu_m}} + \\ &+\frac{\partial}{\partial \boldsymbol{\mu_m}}\operatorname{tr}\left\{\tilde{Y}^\top\tilde{Y} + \underbrace{\sqrt{\boldsymbol{d}^{-1}}^\top\sqrt{\boldsymbol{d}^{-1}}}_{\omega}C\boldsymbol{\mu_m}\boldsymbol{\mu_m}^\top - 2\sqrt{\boldsymbol{d}^{-1}}^\top\tilde{Y}^\top B\boldsymbol{\mu_m}\right\} = 2\omega C\boldsymbol{\mu_m} - 2B^\top\tilde{Y}\sqrt{\boldsymbol{d}^{-1}} - \\ &-\frac{\partial \psi(\boldsymbol{\mu_m})}{\partial \boldsymbol{\mu_m}} = 0 \leftrightarrow \omega C\boldsymbol{\mu_m} - B^\top\tilde{Y}\sqrt{\boldsymbol{d}^{-1}} - \left(A^\top Q_B\right)\odot\left(\mathcal{S}\boldsymbol{1_n}\boldsymbol{1_m}^\top\right)^\top Q_B^\top A\boldsymbol{\mu_m} + A^\top Q_B\left(\mathcal{S}\odot\left(\boldsymbol{r^{-T}}\boldsymbol{1_n}\right)\right)\odot\left(Q_B^\top\hat{Y}\right)\boldsymbol{1_n} = 0 \leftrightarrow \\ &\leftrightarrow \left(\omega C - \left(A^\top Q_B\right)\odot\left(\mathcal{S}\boldsymbol{1_n}\boldsymbol{1_m}^\top\right)^\top Q_B^\top A\right)\boldsymbol{\mu_m} = B^\top\tilde{Y}\sqrt{\boldsymbol{d}^{-1}} - A^\top Q_B\left(\mathcal{S}\odot\boldsymbol{1_n}\boldsymbol{r^{-T}}\right)\odot\left(Q_B^\top\hat{Y}\right)\boldsymbol{1_n},\end{aligned} \tag{119}$$

where $\omega = \sum_{i=1}^n \frac{1}{\sigma_i}$. There is no need to use solvers such as GMRES and exact solution can be obtained as only a matrix of size $m \times m$ has to be inverted. Furthermore, as the model is linear, the inverse of the matrix $\omega C - \left(A^\top Q_B\right)\odot\left(\mathcal{S}\boldsymbol{1_n}\boldsymbol{1_m}^\top\right)Q_B^\top A$ is also an asymptotic covariance matrix of $\boldsymbol{\mu_m}$, i.e.

$$\mathfrak{I}(\boldsymbol{\mu_m}) = \omega C - \left(A^\top Q_B\right)\odot\left(\mathcal{S}\boldsymbol{1_n}\boldsymbol{1_m}^\top\right)^T Q_B^\top A, \tag{120}$$

where $\mathfrak{I}$ is Fisher (expected) information matrix.

# 5. Extending the model to incorporate individual promoter variances

So far we made an assumption of promoter/gene variances, which is unrealistic. Instead, the model should've been

$$\mathring{Y} = \boldsymbol{1_{\mathring{p}}}\ \boldsymbol{\mu_n}^\top + \boldsymbol{\mu_p}\ \boldsymbol{1_n}^\top + \mathring{B}U + \mathring{\mathrm{E}},\ \ \mathrm{E}\sim\mathcal{MN}(0,K,D),\ \ U\sim\mathcal{MN}\left(\boldsymbol{\mu_m}\boldsymbol{1_n}^\top,\Sigma,G\right), \tag{121}$$

where $K$ is a diagonal matrix of promoter/gene variances that should be estimated. All $\mathring{p}$ variance parameters in $K$ can be estimated the REML step where $\mathrm{E}$ is isolated:

$$\underbrace{Q_N^T F_p}_{F}\ \mathring{Y}F_n^T = \hat{\mathrm{E}}, \hat{\mathrm{E}}\sim\mathcal{MN}(0, FKF^T, F_nDF_n^T). \tag{122}$$

Property 4.1.2.2 and Property 4.1.2.3 to the case of general semi-orthogonal matrices $F$, greatly facilitating computation of a loglikelihood for Equation 122.

All the other steps of the algorithm remain unchanged for the most part, as Equation 121 can be transformed to the form very similar to the original Equation 10 just by re-scaling the data:

$$K^{-\frac{1}{2}}\mathring{Y} = \underbrace{\boldsymbol{k}}_{K^{-\frac{1}{2}}\boldsymbol{1_{\mathring{p}}}}\ \ \boldsymbol{\mu_n}^\top + \underbrace{\widehat{\boldsymbol{\mu_p}}}_{K^{-\frac{1}{2}}\boldsymbol{\mu_p}}\ \ \boldsymbol{1_n}^\top + \underbrace{\mathring{\hat{B}}}_{K^{-\frac{1}{2}}\mathring{B}}\ U + \mathring{\mathrm{E}},\ \ \mathrm{E}\sim\mathcal{MN}\left(0,I_{\mathring{p}},D\right),\ \ U\sim\mathcal{MN}\left(\boldsymbol{\mu_m}\boldsymbol{1_n}^\top,\Sigma,G\right). \tag{123}$$

The only change is that the semi-orthogonal centering operator $F_{\mathring{p}}$ should be substituted with a general semi-orthogonal matrix. To this end, we take advantage of the Householder approach scheme introduced in Section 4.4.1. In the principle, this should reduce variance in estimates and improve uncertainty estimation [16] of activities.

This approach is implemented in MARADONER as well when ran with a `--promoter-variance` flag set to the `maradoner fit` command. Some caution must be paid, however: while this approach is theoretically superior, $\mathring{p}$ variance components can be efficiently estimated only when $n$ is large enough. Furthermore, one should pay close attention to filtering the dataset: for instance, presense of a large amount of "silent"/non-expressed genes/promoters in the data will cause the model to downweight active genes/promoters, as it is very appealing for the model to completely explain the large quantity of, effectively, constants, other than trying to explain actually variable genes/promoters.

# 6. Motif clustering

To reduce number of effective parameters and deal with highly correlated motifs, one can apply a clustering on motif loadings in matrix $B$. Here, by "clustering" we mean a non-negative decomposition of the matrix $B = \underbrace{\hat{B}}_{p\times c} \underbrace{L}_{c\times m}$, where $C$ can be interpreted as a matrix of $c$ clusters centers and $L$ is a matrix of loadings onto each of $c$ clusters. Note both $\hat{B}$ and $L$ should contain non-negative elements as the original loadings onto promoters matrix $B$ is non-negative, hence cluster centroids can't be negative, and loadings are non-negative by definition.

## 6.1. General case (NMF)

The NMF (Negative Matrix Factorization) algorithm produces exactly the required decomposition. In the most straightforward case, we can treat the clustering as merely a noise reduction technique in $B$, however it might make final predictions less interpretable and is lacking the performance benefits of a dimensionality reduction. Let's consider a simpler version of Equation 10 with intercepts omited for brevity:

$$Y = \overbrace{\hat{B}L}^{B}\hat{U} + E = \hat{B}\underbrace{U}_{c\times n} + \mathrm{E},\ \mathrm{E} \sim \mathcal{MN}(0, I_p, D),\ U \sim \mathcal{MN}(0, L\Sigma L^\top, G). \tag{124}$$

Some changes should be made to ensure that all previous computations remain correct. Specifically, consider the case when the pushthrough-identity is applied and how it should be adjusted:

$$\begin{aligned} V = \left(\hat{B}L\tilde{\Sigma}\hat{B}^\top L^\top + \sigma I_p\right)^{-1}\hat{B} = \left(\left(\hat{B}L\sqrt{\tilde{\Sigma}}\right)\left(\hat{B}L\sqrt{\tilde{\Sigma}}\right)^\top + \sigma I_p\right)^{-1}\hat{B}L\sqrt{\tilde{\Sigma}}\sqrt{\tilde{\Sigma}}^{-1}\underbrace{\left[L^\top(LL^\top)^{-1}\right]}_{L^+} = \\ = \hat{B}L\sqrt{\tilde{\Sigma}}\left(\left(\hat{B}L\sqrt{\tilde{\Sigma}}\right)^\top\left(\hat{B}L\sqrt{\tilde{\Sigma}}\right) + \sigma I_m\right)^{-1}\sqrt{\tilde{\Sigma}}^{-1}L^+ = B\sqrt{\tilde{\Sigma}}\left(\left(B\sqrt{\tilde{\Sigma}}\right)^\top\left(B\sqrt{\tilde{\Sigma}}\right) + \sigma I_m\right)^{-1}\sqrt{\tilde{\Sigma}}^{-1}L^+ = \\ = A(A^\top A + \sigma I_m)^{-1}\sqrt{\tilde{\Sigma}}^{-1}L^+. \end{aligned} \tag{125}$$

The pseudo-inverse $L^+$ can be computed once before running the optimization procedure. Note that, althogh the dimensionality of the random matrix of motif activities $U$ is indeed reduced from $m$ to $c$, it makes no positive impact on the complexity of computations, making the prospects of applying general NMF dubious here.

## 6.2. Hard clustering

General NMF produces a soft clustering, i.e. the same motif can belong to different clusters. If instead we use a hard clustering algorithm, e.g. K-Means, to produce NMF, the obtained $L$ will have a desirable property: $L$ is orthogonal. i.e. $LL^\top = D_L$, where $D_L$ is a diagonal matrix. It follows that for a given motif $k$ there is a single cluster $j$ it belongs to, i.e. $L_{j,k} = 1$ and $L_{i,k} = 0\ \forall i \neq j$.

Therefore,

$$L\underbrace{\Sigma}_{m\times m}L^\top = \underbrace{\hat{\Sigma}}_{c\times c}, \tag{126}$$

where $\hat{\Sigma}$ is diagonal too. More specifically, each diagonal term is a sum of $\tau_k$ variances of all motifs that belong to a given cluster. It has 2 practical implications:

1. It means that if we use hard clustering algorithm and fully parameterized diagonal $\Sigma$, we can just substitute $B$ with $\hat{B}$ and call it a day;
2. If we are not interested in estimating motif variances, i.e. when $\Sigma = I_m$, then we substitute $B$ with $\hat{B}$ with each of its $c$ columns scaled by a reciprocal of a root of a number of elements in a particular cluster.

In both cases, the complexity is reduced everywhere from $m$ to $c$. We conclude that hard-clustering, although less versatile in terms of a, well, clustering, brings significantly more computational benefits.

# 7. Motif activity prediction

## 7.1. Maximum-a-posteriori estimates

For obtaining $U$ estimates, we use maximum-a-posteriori (MAP) approach:

$$U_{\mathcal{MAP}} = \underset{U}{\operatorname{argmax}}\, f(Y, U|G, D, \Sigma) = \underset{U}{\operatorname{argmax}} \ln f(Y, U|G, D, \Sigma) = \underset{U}{\operatorname{argmax}}[\ln f(Y|U, D) + \ln f(U|G, \Sigma)], \tag{127}$$

where $f$ is a likelihood function. The conditional distribution is defined by the mode in Equation 10, as well as the marginal distribution of $U$. It is more practical to rewrite Equation 127 for some particular group as they are independently modeled (although linked together with the same $\Sigma$):

$$u_{\mathcal{MAP}} = \underset{u}{\operatorname{argmax}}[\ln f(y|u, \sigma) + \ln f(u|\nu, \Sigma)] = \underset{u}{\operatorname{argmax}}[-F(u)] \tag{128}$$

Let's expand the function $F(u)$ while canceling out terms that do not depend on $u$:

$$\begin{aligned} F(u) &= u^\top \tilde{\Sigma}^{-1} u + \sum_{i=1}^{\hat{n}} \frac{1}{\sigma}(Y_i - Bu)^\top (Y_i - Bu) = u^\top \tilde{\Sigma}^{-1} u + \frac{1}{\sigma} \sum_{i=1}^{\hat{n}} \cancel{Y_i^\top Y_i} - Y_i^\top Bu - u^\top B^\top Y_i + u^\top B^\top B u \rightarrow \\ \rightarrow F(u) &= u^\top \tilde{\Sigma}^{-1} u + \frac{1}{\sigma} \sum_{i=1}^{\hat{n}} u^\top B^\top B u - 2u^\top B^\top Y_i = u^\top \tilde{\Sigma}^{-1} u + \frac{\hat{n}}{\sigma}\left(u^\top B^\top B u - 2u^\top B^\top \overline{Y}\right) = \\ &= u^\top \left[\tilde{\Sigma}^{-1} + \frac{\hat{n}}{\sigma} B^\top B\right] u - 2\frac{\hat{n}}{\sigma} u^\top B^\top \overline{Y}, \end{aligned} \tag{129}$$

where $\underbrace{\overline{Y}}_{p\times 1}$ is a sample average of expression observations for a particular group. Let's compute Cholesky decomposition $Z^{-1} = \tilde{\Sigma}^{-1} + \frac{\hat{n}}{\sigma} B^\top B = L^\top L$, then:

$$\begin{aligned} F(u) &= u^\top L^\top L u - 2\frac{\hat{n}}{\sigma} u^\top B^\top \overline{Y} = u^\top L^\top L u - 2\frac{\hat{n}}{\sigma} u^\top L^T L^{-T} B^\top \overline{Y} \pm \frac{\hat{n}}{\sigma} \overline{Y}^\top B L^{-1} L^{-T} B^\top \overline{Y} = \\ &= \left(Lu - \frac{\hat{n}}{\sigma} L^{-T} B^\top \overline{Y}\right)^\top \left(Lu - \frac{\hat{n}}{\sigma} L^{-T} B^\top \overline{Y}\right) - \cancel{\frac{\hat{n}}{\sigma} \overline{Y}^\top B L^{-1} L^{-T} B^\top \overline{Y}} = \\ &= \left(u - \frac{\hat{n}}{\sigma} L^{-1} L^{-T} B^\top \overline{Y}\right)^\top L^\top L \left(u - \frac{\hat{n}}{\sigma} L^{-1} L^{-T} B^\top \overline{Y}\right) = \left(u - \frac{\hat{n}}{\sigma} Z B^\top \overline{Y}\right)^\top Z^{-1} \left(u - \frac{\hat{n}}{\sigma} Z B^\top \overline{Y}\right). \end{aligned} \tag{130}$$

This form coincides with the multivariate normal distribution. It can be said that posterior distribution of $u$ given sample data for a particular group is:

$$u \sim \mathcal{N}\left(\frac{\hat{n}}{\sigma} Z B^\top \overline{Y}, Z\right), \quad \text{where} \quad Z = \left(\tilde{\Sigma}^{-1} + \frac{\hat{n}}{\sigma} B^\top B\right)^{-1}, \; \tilde{\Sigma} = \nu\Sigma. \tag{131}$$

Then, the maximization problem in Equation 128 is solved with

$$u_{\mathcal{MAP}} = E[u] = \frac{\hat{n}}{\sigma} Z B^\top \overline{Y}. \tag{132}$$

## 7.2. Improving estimates with cross-validation (CV)

It is possible to slightly enhance activities estimates if we have some extra time at our disposal at the prediction stage (usually the case, MAP estimates are fast to compute given parameters set). Let's take a look at the MAP estimates $E[u]$:

$$\begin{aligned} E[u] &= \frac{\hat{n}}{\sigma} Z B^\top \overline{Y} = \frac{\hat{n}}{\sigma} \left(\underbrace{\tilde{\Sigma}^{-1}}_{\frac{1}{\nu}\Sigma^{-1}} + \frac{\hat{n}}{\sigma} B^\top B\right)^{-1} B^\top \overline{Y} = \frac{\hat{n}}{\sigma} \left(\frac{1}{\nu}\Sigma^{-1} + \frac{\hat{n}}{\sigma} B^\top B\right)^{-1} B^\top \overline{Y} = \\ &= \hat{n} \left(\underbrace{\frac{\sigma}{\nu}}_{\mu^{-1}} \Sigma^{-1} + \hat{n} B^\top B\right)^{-1} B^\top \overline{Y} = \hat{n} \left(\frac{1}{\mu}\Sigma^{-1} + \hat{n} B^\top B\right)^{-1} B^\top \overline{Y}. \end{aligned} \tag{133}$$

Here, $\mu$ can be thought of as a "signal to noise ratio". Higher the $\mu$, the greater part of variance is explained by the model. In the limit, as $\mu \rightarrow \infty$, the model degenerates into an ordinary linear regression with a pseudo-inverse/mean-square solution, at the risk of an overfitting, of course.

$\mu$ is then can be treated as a hyperparameter and tuned using CV. After an optimal value of $\mu$ is found, we deduce the new tuned $\nu$ from it as $\nu = \mu\sigma$, usable later on in the covariance matrix $Z$.

# 8. Tests

There are two ways to do hypothesis testing in MARADONER. The first one takes advantage of asymptotic properties of parameter estimates (namely, $\Sigma$ and $\boldsymbol{\mu_m}$). The second one uses posterior distribution of motif activities $U$ as obtained in the previous section.

## 8.1. Asymptotic approach

It's well-known that maximum-likelihood estimates are asymptotically distributed as multivariate-normal with the covariance matrix equal to the inverse of the Fisher information matrix $\mathfrak{I}$, as stated by the Rao-Cramer bound. This allows us to test for hypotheses such as:

1. Does any particular motif change its activity across different groups? To this end, we test a hypothesis $\tau_i > 0$;
2. Is a motif, at average, a repressor or an enhancer? That's done by checking that a particular element of $\boldsymbol{\mu_m} > 0$ or $< 0$. In practice, it is very uncommon for a motif to be neither.

## 8.2. Using posterior distribution of $U$

On the other hand, we can test a hypothesis by analysing the posterior distribution of $U$. That's the original MARA approach as well.

### 8.2.1. Hypothesis testing

After "studentization" of the $U$ MAP estimates, we run different tests:

1. Z-test: is the particular motif active in this particular group?
2. ANOVA: is this particular motif active anywhere? Implemented via $\chi^2(g-1)$ by comparing each of the group-wise $U$ predictions against the BLUE
$$\mu_u = \left(\sum_{i=1}^{g} \Sigma_i^{-1}\right)^{-1} \left(\sum_{i=1}^{g} \Sigma_{i=1}^{-1} U_{.i}\right), \tag{134}$$
where $\Sigma_i$ is the posterior covariance matrix of $U_{.i}$.
3. "Off-test": Is there at least one group where the motif is not actve? Implemented via examining the distribution of minimums of squared motif activities.

# 9. Algorithm overview

$$\mathring{Y} = \boldsymbol{1_{\mathring{p}}} \boldsymbol{\mu_n}^\top + \boldsymbol{\mu_p} \boldsymbol{1_n}^\top + \mathring{B}U + \mathring{\mathrm{E}},\ \mathrm{E} \sim \mathcal{MN}\big(0, I_{\mathring{p}}, D\big),\ U \sim \mathcal{MN}\big(\boldsymbol{\mu_m} \boldsymbol{1_n}^\top, \Sigma, G\big), \tag{135}$$

1. Kill off nuisance parameter $\boldsymbol{\mu_n}$ by using the orthogonal centering operator $H_n$ on $\mathring{Y}$:
$$H_{\mathring{p}}\mathring{Y} = \cancel{H_{\mathring{p}} \boldsymbol{1_{\mathring{p}}}} \boldsymbol{\mu_n}^\top + H_{\mathring{p}} \boldsymbol{\mu_p} \boldsymbol{1_n}^\top + H_{\mathring{p}} BU + H_{\mathring{p}} \mathring{\mathrm{E}}. \tag{136}$$
2. Kill off both $\boldsymbol{\mu_p}$ $\boldsymbol{\mu_m}$ by transforming $Y$ by $H_n^\top$ and then also kill-off $B$ by transforming $Y$ by $Q_N$ to estimate just $D$ alone :
$$Q_N^\top H_{\mathring{p}} \mathring{Y} H_n^\top = Q_N^\top H_{\mathring{p}} \boldsymbol{\mu_p} \cancel{\boldsymbol{1_n}^\top H_n^\top} + \cancel{Q_N^\top H_{\mathring{p}} \mathring{B}} U H_n^\top + Q_N^\top H_{\mathring{p}} \mathring{\mathrm{E}} H_n^\top . \tag{137}$$
3. Get back $B$, but keep transforming by $H_n$ to keep $\boldsymbol{\mu_m}$ and $\boldsymbol{\mu_p}$ away, estimate $\Sigma$ and $G$:
$$H_{\mathring{p}} \mathring{Y} H_n^\top = H_{\mathring{p}} \boldsymbol{\mu_p} \cancel{\boldsymbol{1_n}^\top H_n^\top} + H_{\mathring{p}} \mathring{B} \hat{U} + H_{\mathring{p}} \mathring{\mathrm{E}} H_n^\top,\ \hat{U} \sim \mathcal{MN}\big(\boldsymbol{\mu_m} \cancel{\boldsymbol{1_n}^\top H_n^\top}, \Sigma, H_n G H_n^\top\big) \tag{138}$$
4. (This step can be done before estimating $\Sigma$ and $D$ as well) Re-introduce $Q_N$ to zero out $B$, estimate $\boldsymbol{\mu_p}$:
$$Q_N^\top H_p \mathring{Y} = Q_N^\top H_{\mathring{p}} \boldsymbol{\mu_p} \boldsymbol{1_n}^\top + \cancel{Q_N^\top H_{\mathring{p}} \mathring{B}} U + Q_N^\top H_p \mathrm{E}. \tag{139}$$
5. Subtract promoter-wise means $\boldsymbol{\mu_p}$ and given all other estimates $\Sigma, G$ and $D$ estimate motif-wise means $\boldsymbol{\mu_m}$:
$$H_{\mathring{p}} \mathring{Y} - H_{\mathring{p}} \boldsymbol{\mu_p} \boldsymbol{1_n}^\top = H_{\mathring{p}} \mathring{B} U + Q_N^\top H_{\mathring{p}} \mathring{\mathrm{E}}. \tag{140}$$
6. Obtain MAP estimates of $U$.
7. Given both MAP estimates and the asymptotic covariance matrices of parameter estimates, do statistical tests.

## 10. Gene Regulatory Network (GRN) extension

MARADONER has a built-in heuristic tool to infer which promoters are being affected by a motif in a particular group. To this end, parameter estimates are obtained, as explained in previous sections. Then, for an $i$-th promotor/gene in $j$-th group we fit the following linear model

$$\mathring{Y}_{i\mathfrak{g}(j)} = \boldsymbol{\mu}_{\boldsymbol{p}_i}\mathbf{1_s}^T + \boldsymbol{\mu}_{\boldsymbol{s}_{\mathfrak{g}(j)}}\mathring{B}_{i\cdot}U_{\cdot\mathfrak{g}(j)} + \varepsilon,\ U \sim \mathcal{MN}\left(\boldsymbol{\mu}_{\boldsymbol{m}}\mathbf{1_n}^\top, \Sigma, G\right),\ \varepsilon \sim \mathcal{N}\left(0, \delta_i^2 I_{N(j)}\right), \tag{141}$$

where $\mathfrak{g}(j)$ returns a set of indices corresponding to samples of the $j$-th group. The only quantity estimated here is $\delta_i^2$ – promoter- and group-specific error term variance. Then, for a $k$-th motif, we compure Bayesian belief [17], comparing likelihoods of $\mathcal{H}_0$ (Equation 141) against $\mathcal{H}_1$, constructed as Equation 141 with $k$-th motif effects deleted (it is assumed that deletion of a single motif causes no significant change in $\delta_i^2$):

$$P\left(\mathcal{H}_0|\mathring{Y}_{i\mathfrak{g}(j)}\right) = \frac{L\left(\mathring{Y}_{i\mathfrak{g}(j)}|\mathcal{H}_0\right)}{L\left(\mathring{Y}_{i\mathfrak{g}(j)}|\mathcal{H}_0\right)P(\mathcal{H}_0) + L\left(\mathring{Y}_{i\mathfrak{g}(j)}|\mathcal{H}_1\right)(1 - P(\mathcal{H}_0))}, \tag{142}$$

where $L(\cdot)$ denotes a likelihood for a particular hypothesis, $P(\mathcal{H}_0)$ is a prior probability of a motif having an effect on a promoter. By default, MARADONER uses a neutral value of $P(\mathcal{H}_0) = \frac{1}{2}$.

## 11. Simulation studies

In order to assess the performance of MARADONER in a variety of settings, a synthetic data generator was developed. In addition, the MARA approach was re-implemented in order to establish a robust baseline for the purposes of comparison.

The data generator samples expression matrix $Y$ for a given set of parameters $K, D, G, \Sigma, G, \mu_p, \mu_s, \mu_m$ and a randomly generated loading matrix $B$, following Equation 121. The generator has the following parameters:

1. Parameters related to a dataset size: a number of genes $p$, a number of samples $s$, a number of motifs $m$;
2. Two parameters controlling error term $E$ heteroscedasticity across genes dimension: S_het – if False, $K = I_p$, and S_var_max, controlling the maximal variance component. Each component of $K$ is sampled from $\mathcal{U}(0.1, \text{S_var_max})$;
3. Two parameters controlling heteroscedasticity of $U$: Sigma_het – if False, $\Sigma = I_m$, and Sigma_var, controlling the spread of $\Sigma$ values. Each component of $K$ is sampled from $\mathcal{LN}(0, \text{Sigma_var})$, where $\mathcal{LN}$ is Lognormal;
4. Fraction of motif activities to have no effect (and, consequently, zero variances in $\Sigma$): zm_frac.
5. Expected fraction of variance explained by a random effects matrix $U$: variance_ratio.

As for $D$ and $G$, they are sampled from $\mathcal{U}(1.0, 2.5)$ and $\mathcal{U}(0.1, 2.0)$ respectively, but $D$ is rescaled appropriately to comply with the variance_ratio parameter. Intercepts are sampled from $\mathcal{N}(0, 1)$, elements of $B$ are sampled from $\mathcal{U}(0.1, 1.1)$.

In total, we generated $39$ datasets of various configurations, as depicted at Table 2, each one was regenerated with a different rng seed $20$ times to compute reliable metrics, making $780$ datasets in total. We found following metrics worth inspecting:

1. Pearson correlation coefficient (PCC) between predicted "expression" $Y$ values and the ground truth. To this end, we hold of $10\%$ of "genes" as a testing dataset;
2. PCC between the estimated $U$ and the ground truth;
3. Mean absolute percentage error (MAPE) for parameters in $D$ and $G$. Note that if $\Sigma$ or $K$ are estimated, those parameters become unidentifiable up to a scale, hence we have to rescale estimates so that MAPE makes sense.

Mean values of those metrics, averaged across 20 samples, are shown at Figure 1. Note that for groups where S_het = True the notion of MAPE for $D$ and $G$ loses its interpretability as a goodness-of-estimation measure due to the distortions in the variance structure.

For the dataset group $F$ we also show MAPE plot for promoter/gene variances $K$ as a function of a number of samples, as it bears a great practical interest – see Figure 2.

So far we have examined datasets that were generated in an agreement with the MARADONER model. However, although MARADONER attempts to capture a greater degree of variability in data than conventional MARA-like approaches, it still falls short to acknowledge the most apparent discrepancy between the model and the real experimental data: genetic variances are not the same across all groups and samples, but must differ as well. From a modelling standpoint, acknowledging this would lead to a necessity of introducing group-specific gene-wise variances, which might not be feasible from a computational standpoint, nor feasible from a standpoint of statistical estimation, as the number of samples per group is scarce. Therefore, it is interesting to see MARADONER (with estimable global gene-wise variance matrix $K$) behaves under such a discrepancy. To this end, we added two extra parameters to the data generator: S_het_sample and S_sample_var, controlling the degree of variability in the data generator. If S_het_sample is True, then it works as follows: first, as before, an expected average variance is sampled for the $k$-th gene $\hat{s}_k \sim \mathcal{U}(0.1, \text{S_max})$; then the sample-specific variance is sampled from $\mathcal{LN}(\hat{s}_k, \text{S_sample_var})$. Configurations of those "super" heterscedastic datasets are shown at Table 3. MARADONER with estimable global gene-wise variances appears to be rather robust to such a specification error, as evident by the Figure 3.

Table 2: Configurations of generated datasets.

| | $m$ | $p$ | $s$ | variance_ratio | zm_frac | Sigma_het | Sigma_var | S_het | S_var_max |
|---|---|---|---|---|---|---|---|---|---|
| A | 100 | **1000, 2000, 4000, 5000, 8000, 10000, 20000** | 20 | 0.1 | 0.0 | False | - | False | - |
| B | 100 | 5000 | **2, 4, 8, 16, 20, 32, 64, 128** | 0.1 | 0.0 | False | - | False | - |
| C | 100 | 5000 | 20 | **0.05, 0.1, 0.2, 0.3** | 0.0 | False | - | False | - |
| D | 100 | 5000 | 20 | 0.1 | **0.0, 0.05, 0.1, 0.2, 0.3, 0.4** | False | - | False | - |
| E | 100 | 5000 | 20 | 0.1 | 0.0 | **True** | **0.1, 0.5, 1, 2, 4, 10, 32** | False | - |
| F | 100 | 5000 | 20 | 0.1 | 0.0 | False | - | **True** | **1.5, 2, 4, 8, 16, 32** |
| G | 100 | 5000 | **8, 16, 32, 64, 128, 256** | 0.1 | 0.0 | False | - | **True** | 2 |

Table 3: Configurations of generated datasets with sample-specific gene-wise variances. Other params are $p = 5000$, $m = 100$, S_var_max = 2.5, variance_ratio = 0.1.

| | $s$ | S_sample_var | Sigma_het | Sigma_var | zm_frac |
|---|---|---|---|---|---|
| H | **8, 16, 32, 48, 64, 128, 256** | 0.1 | False | - | 0 |
| I | 32 | **0.1, 0.15, 0.2, 0.4, 1.0** | False | - | 0 |
| J | 128 | **0.1, 0.15, 0.2, 0.4, 1.0** | False | - | 0 |
| K | **8, 16, 32, 48, 64, 128, 256** | 0.2 | True | 4 | 0.2 |

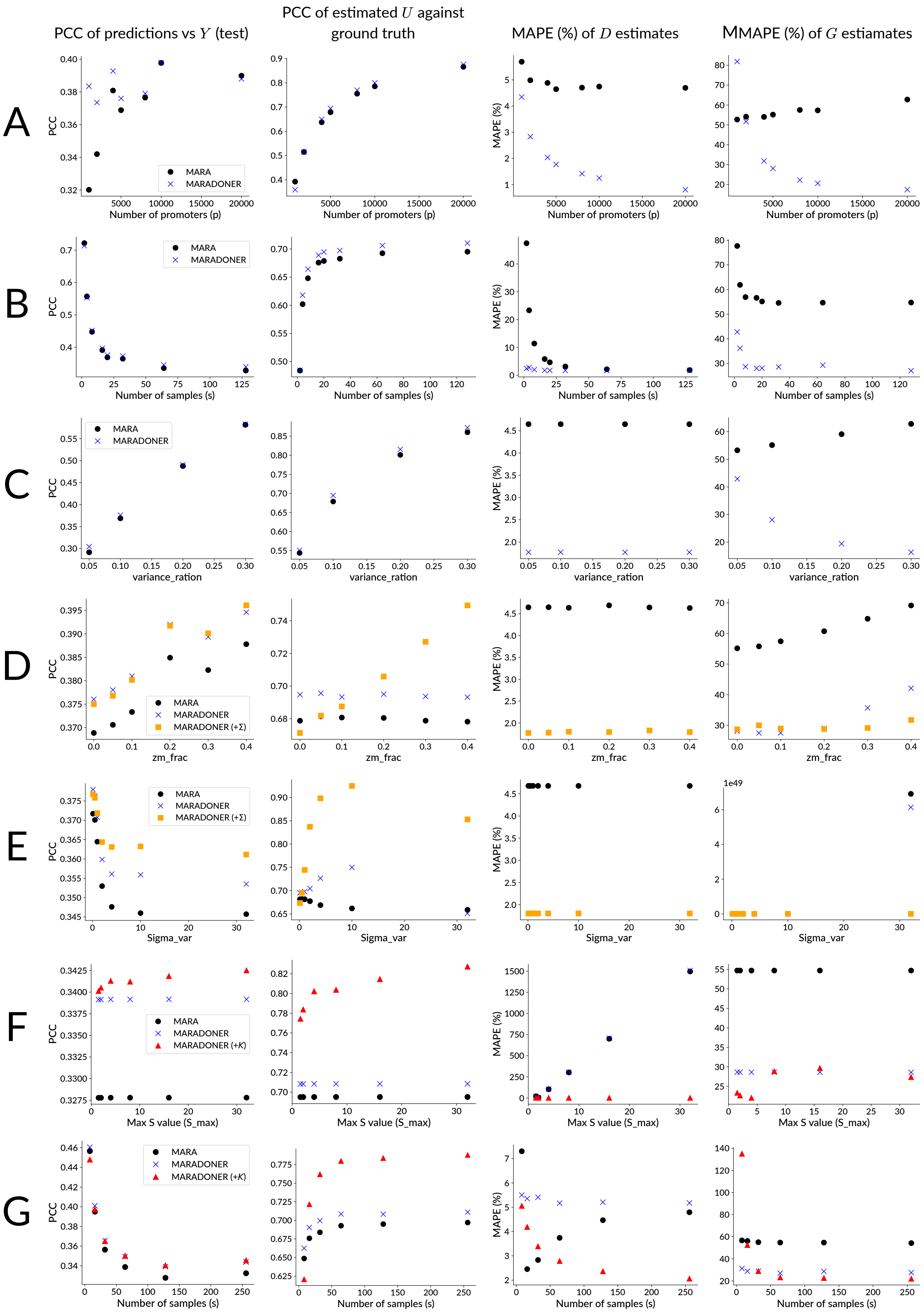

Figure 1: Evaluated metrics for datasets A-G.

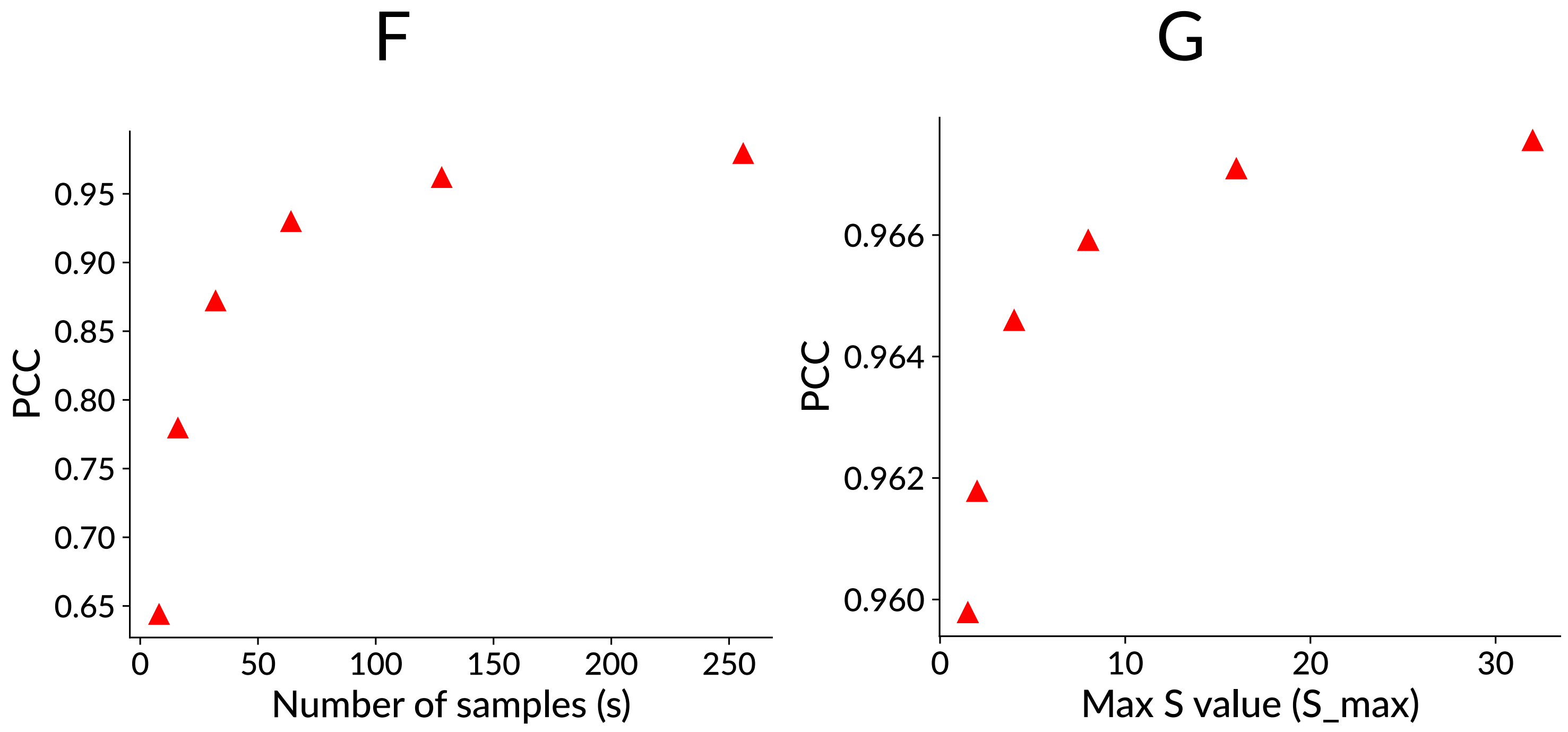

Figure 2: Pearson correlation coefficient between ground truth and estimated promoter variances for dataset groups F and G.

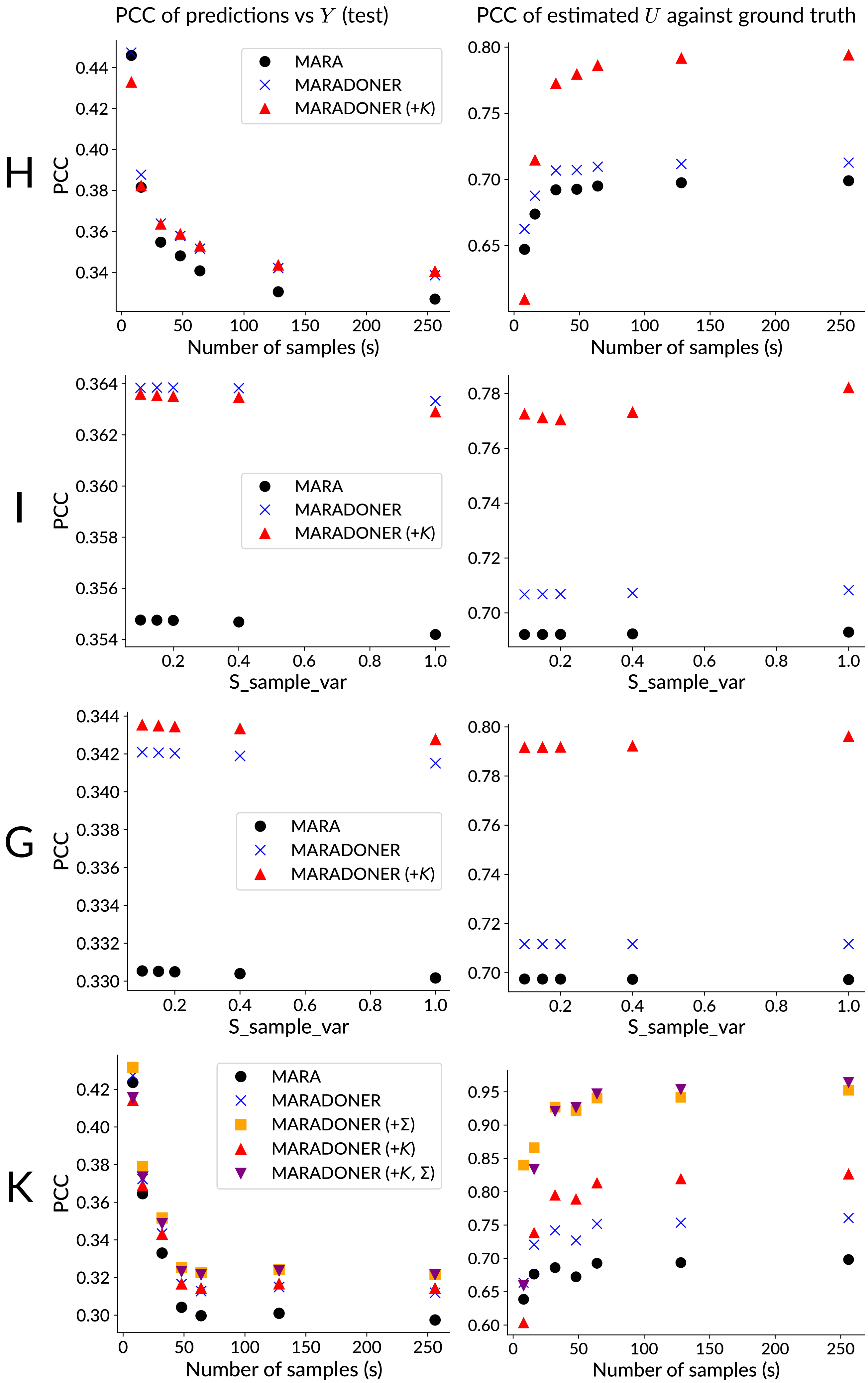

Figure 3: Evaluated metrics for datasets H-K that violate variance assumptions of MARADDONER.